\documentstyle[12pt] {article}
\def\beq{\begin{equation}}
\def\eeq{\end{equation}}
\def\bea{\begin{eqnarray}}
\def\eea{\end{eqnarray}}
\def\bq{\begin{quote}}
\def\eq{\end{quote}}

\begin{document}
\pagestyle{empty}
\begin{flushright}
\end{flushright}
\vspace*{5mm}
\begin{center}
{\bf  QUASI $BPS$ WILSON LOOPS, LOCALIZATION OF LOOP EQUATION BY
 HOMOLOGY AND EXACT BETA
 FUNCTION IN THE LARGE-$N$ LIMIT OF $SU(N)$ YANG-MILLS THEORY}
\\
\vspace*{1cm}
{\bf M. Bochicchio} \\
\vspace*{0.5cm}
INFN Sezione di Roma \\
Dipartimento di Fisica, Universita' di Roma `La Sapienza' \\
Piazzale Aldo Moro 2 , 00185 Roma  \\
e-mail: marco.bochicchio@roma1.infn.it \\
\vspace*{2cm}
{\bf ABSTRACT  } \\
\end{center}
\vspace*{5mm}
\noindent
 We localize the loop equation of large-$N$ $YM$ theory in the $ASD$ variables on a 
 critical equation for an effective action by means of homological methods as opposed to
 the cohomological localization of equivariantly closed forms in local field theory.
 Our localization occurs for some special simple quasi $BPS$ Wilson loops, 
 that have no perimeter divergence and no cusp anomaly for backtracking cusps,
 in a partial Eguchi-Kawai reduction from four to two dimensions of the non-commutative theory 
 in the limit of infinite non-commutativity and in a lattice regularization in which
 the $ASD$ integration variables live at the points of the lattice, thus implying an embedding
 of parabolic Higgs bundles in the $YM$ functional integral.
 Homological localization is based on an analogy with cohomological localization.
 The analog of the invariance of the cohomological class of a closed form
 for the addition of a co-boundary is
 the zig-zag symmetry, i.e. the invariance of the holonomy class of a quasi $BPS$ Wilson loop
 for the addition of the boundary of a tiny strip, of size of the cut-off.
 The analog of the action being a closed form 
 is the invariance of the $RG$ flow of the Wilsonean renormalized effective action in the $ASD$ variables,
 $\Gamma_q$, under the local conformal
 transformation that generates the holonomically trivial deformation of the quasi
 $BPS$ Wilson loops.
 Finally, the analog of the rescaling by a divergent factor of the cohomological trivial
 deformation that localizes the exponential of a closed form is the local conformal
 rescaling that maps any (lattice) marked point of a 
 quasi $BPS$ Wilson loop to a cusp at infinity. 
 The homological reason for localization is that the contact term that occurs
 at the marked points of the loop, that is the obstruction to localization in the loop equation, 
 vanishes at the cusps because of the absence of cusp anomaly for the backtracking cusps of the quasi $BPS$
 Wilson loop and the manifest zig-zag invariant regularization of the loop
 equation in the $ASD$ variables.
 Yet, a posteriori, we find a simple reason for homological localization to occur:
 $\Gamma_q$ flows to the ultraviolet by the localizing conformal transformation
 and thus, since it is $AF$, to vanishing coupling.
 We find that the beta function of $\Gamma_q$ is saturated by the non-commutative $ASD$ vortices of 
 the $EK$ reduction. An exact canonical beta function of $NSVZ$ type
 that reproduces the universal first and second perturbative coefficients
 follows by the localization of $\Gamma_q$ on vortices.
 Finally we argue that a scheme can be found in which the canonical coupling coincides
 with the physical charge between static quark sources in the large-$N$ limit 
 and we compare our theoretical calculation with some numerical lattice result.
  
\vspace*{1cm}
\begin{flushleft}
\end{flushleft}
\phantom{ }
\vfill
\eject
\setcounter{page}{1}
\pagestyle{plain}

\section{Introduction}

In this paper we revisit and refine our result \cite{MB1} on the exact beta function
in the large-$N$ limit of the pure Yang-Mills theory.
From a computational point of view we can summarize our result as follows. \par
There exists a renormalization scheme in which the large-$N$ canonical beta function of the pure $YM$ theory is given by:
\bea
\frac{\partial g_c}{\partial log \Lambda}=\frac{-\beta_0 g_c^3+
\frac {\beta_J}{4} g_c^3 \frac{\partial log Z}{\partial log \Lambda} }{1- \beta_J g_c^2 }
\eea
with:
\bea
\beta_0=\frac{1}{(4\pi)^2} \frac{11}{3} \nonumber \\
\beta_J=\frac{4}{(4\pi)^2}
\eea
where $g_c$ is the 't Hooft canonical coupling constant and $ \frac{\partial log Z}{\partial log \Lambda} $ 
is computed to all orders in the 't Hooft Wilsonean coupling constant, $g_W$, by:
\bea
\frac{\partial log Z}{\partial log \Lambda} =\frac{ \frac{1}{(4\pi)^2} \frac{10}{3} g_W^2}{1+cg_W^2}
\eea
with $c$ a scheme dependent arbitrary constant.
At the same time, the beta function for the 't Hooft Wilsonean coupling
is exactly one loop:
\bea
\frac{\partial g_W}{\partial log \Lambda}=-\beta_0 g_{W}^3
\eea
Once
the result for $ \frac{\partial log Z}{\partial log \Lambda} $ to the lowest order in the canonical
coupling 
\bea
\frac{\partial log Z}{\partial log \Lambda}=
\frac{1}{(4\pi)^2} \frac{10}{3} g_c^2 + ...
\eea
is inserted in Eq.(1), it implies the correct value of the first and
second perturbative coefficients of the beta function \cite{AF,AF1,2loop1,2loop2}:
\bea
\frac{\partial g_c}{\partial log \Lambda}=
-\beta_0 g_c^3+
(\frac {\beta_J}{4} \frac{1}{(4\pi)^2} \frac{10}{3} -\beta_0 \beta_J) g_c^5 +... \nonumber \\
=-\frac{1}{(4\pi)^2}\frac{11}{3} g_c^3 + \frac{1}{(4\pi)^4} ( \frac{10}{3}
-\frac{44}{3})g_c^5 +... \nonumber \\
=-\frac{1}{(4 \pi)^2} \frac{11}{3} g_c^3 -\frac{1}{(4 \pi)^4} \frac{34}{3} g_c^5+...
\eea
which are known to be universal, i.e. scheme independent. \par
We present in this paper a more general and simplified computation of the effective action that leads to the 
exact beta function. \par
In addition we argue that there is a scheme in which the canonical coupling coincides with
a certain definition of the physical effective charge in the inter-quark potential.
In this scheme the beta function is given by:
\bea
\frac{\partial g_{phys}}{\partial log r}=\beta_0 \frac{g_{phys}^3}{1- \beta_J g_{phys}^2 }
\frac{log(r\Lambda_W)}{log(r\Lambda_W)- \frac{15}{121}} 
\eea
with $\Lambda_W$ the $RG$ invariant scale in the Wilsonean scheme. \par
But overall we furnish a mathematical and a physical explanation for the exactness of the
beta function. \par
On the mathematical side our result for the beta function is based on localizing
on a saddle-point equation for an effective action a version of the large-$N$ loop equation for
certain quasi $BPS$ Wilson loops,
in which the functional integration is, by a change of variables, on the anti-self-dual ($ASD$)
part of the curvature of the connection. \par
The mentioned quasi $BPS$ Wilson loops of the large-$N$ pure $YM$ 
theory enjoy non-renormalization properties analogous to the ones of $BPS$ Wilson loops
of theories with extended supersymmetry \cite{Gr}. 
In particular we find that they do not have perimeter and cusp divergences for 
backtracking cusps. \par
We may wonder how this localization can be possible. 
In fact there are examples of non-trivial quantum field theories in which some special observables
can be evaluated exactly at a saddle-point.
The rationale behind these examples is a version of
the localization described in \cite{DE}, introduced in quantum field theories in \cite{W1}.
Some of the most interesting cases are the cohomology ring of $d=2$ $YM$ theory \cite{W1},
the prepotential of $d=4$ $\cal{N}$ $=2$ $SUSY$ gauge theory \cite{N} and more recently
some circular Wilson loops in $d=4$ $\cal{N}$ $=4,2$ $SUSY$ gauge $YM$ theories \cite{P}.
Localization in quantum field theory is based on cohomological concepts \cite{DE,W1} that involve the action of
a nilpotent $BRST$ differential on equivariantly closed forms. The proof of localization is obtained
noticing that we can add freely to an exponentiated equivariantly closed form 
the $BRST$  variation of any globally defined form without changing its integral.
Then the added $BRST$ differential is rescaled by a divergent factor to get localization
by a semi-classical argument. \par
In all the mentioned four-dimensional cases the $BRST$ charge on which localization is based 
is the supercharge of a kind
of twisted supersymmetry of the action and of the observables. Thus localization, when it applies, applies 
only to special observables
and cannot be extended to all the observables of the theory. \par
The localization
in the large-$N$ pure $YM$ theory that we refer to is furnished by a certain version of the large-$N$ loop equation
and it is not based at all on a form of twisted $SUSY$, that of course is absent in the $YM$
theory. \par 
Rather, and perhaps not surprisingly, the localization of the loop equation on a saddle point for
an effective action becomes possible only
after exploiting a stringy character of the loop equation for the quasi $BPS$ Wilson loops
in which the zig-zag symmetry \cite{Pol} of the Wilson loops plays a key role. \par
The zig-zag symmetry is the invariance of our quasi $BPS$ Wilson loops for the addition of a backtracking arc.
By means of the zig-zag symmetry we can add freely to the Wilson loop backtracking arcs
of arbitrary length that end into cusps.
In a regularized version we can substitute the backtracking arc with the boundary of a tiny strip,
of size of the cut-off. \par
Thus the localization that we describe in the loop equation is a homological concept,
rather than cohomological, and its proof is therefore obtained drawing a certain (weighted) 
arc family on the loop of the loop equation, that at topological level obeys the axioms of 
topological closed/open string theory \cite{Pen}. \par
It is of the utmost importance that the zig-zag symmetry be explicitly non-anomalous, i.e. that it
holds without fine-tuning of the renormalization scheme,
for localization to apply in our version of the loop equation.
This is the case for the quasi $BPS$ Wilson loops that we refer to, because of the absence of
cusp anomaly for backtracking cusps and of perimeter divergence.
Since the zig-zag symmetry is the homological equivalent of the cohomological invariance
of the cohomology class of a closed form by the addition
of the $BRST$ differential of a globally defined form,
alone it is by far too general to suffice to localize the loop equation,
as it is its corresponding notion in cohomology. 
Indeed we will see that there exist renormalization schemes in which the zig-zag symmetry
holds for all the usual unitary Wilson loops, that form a complete set of the theory.
It is very unlikely that any form of localization will ever apply to all these observables. \footnote{Formally the computations
of this paper apply, mutatis mutandis, also to some diagonally embedded Wilson loops
whose curvature is, following our conventions, of $SD$ type. The associated connection is, in the notation
of sect.2, $B^{phys}=(A_z+D_u) dz+(A_{\bar z}-D_{\bar u}) d \bar z$ and thus it is hermitean.
These Wilson loops are physical in the sense
that they have perimeter divergence and cusp anomaly and they imply a non-trivial inter-quark potential.
Thus, for them, the zig-zag symmetry holds only
by fine-tuning of the renormalization scheme. In addition, for them, the $EK$ reduction does not lead
to the parabolic Higgs bundles of Hitchin type of sect.3 .} \par
Another, more dynamical, crucial ingredient is necessary.
In fact, following the analogy with cohomological localization, we need the analog of the property
that the action of the theory is an equivariantly closed form. 
We can rephrase this property by saying that there is a $BRST$ symmetry of the action that generates
the cohomologically trivial deformations.
In our homological language there is no such symmetry of the action, but there is a symmetry
of the $RG$ flow of the Wilsonean renormalized effective action. 
In fact for the quasi $BPS$  Wilson loops
the two-dimensional local conformal transformation that generates the holonomically trivial deformation
of the loop \cite{Pen1}
can be lifted to a local conformal transformation of the four-dimensional theory, because of the quasi
$BPS$ constraint (see sect.2 and sect.3).
The essential reason for this lifting is that a quasi $BPS$ Wilson 
loop lives on a two-dimensional Riemann surface
$\Sigma$ that is diagonally embedded in the four-dimensional theory. \par
This conformal transformation moves the Wilsonean renormalized action along the $RG$ trajectory since
it is equivalent to add to it, as a local counter-term, the conformal anomaly.
The conformal anomaly is computed exactly
a posteriori from the exact beta function of the localized effective action. \par
From this point of view the reason a posteriori for homological localization is that the
Wilsonean renormalized effective action flows to the ultraviolet by the local conformal map and thus, being $AF$,
to vanishing coupling. Then a semi-classical argument implies localization
as in the cohomological case.\par
However we can find a purely homological reason for localization to apply in the
loop equation in the $ASD$ variables. As we have anticipated, it turns out that the loop equation can be localized 
on a critical equation only on
a weighted graph \cite{Pen} whose spine, obtained collapsing all the tiny strips to strings,
is a Mandelstam graph \cite{Man} on which backtracking strings that end into cusps are drawn.
On such a graph each string meets transversely the Wilson loop.
In addition each string starts and ends into a cusp in such a way that the Wilson loop
separates the cusps pairwise into two regions.
The effective action on the weighted graph has a stringy character in the sense that
its local part
is in fact bi-local, since it is localized on pairs of cusps having a string in common.
The homological reason for which the loop equation in the $ASD$ variables localizes on the weighted graph
is that the usual obstruction to localization in the loop equation,
that is the contact term, vanishes at the backtracking cusps of the Mandelstam graph,
because of the zig-zag symmetry in our (regularized) version of the loop equation and because 
of the absence of cusp anomaly for the
quasi $BPS$ Wilson loop. \par
The cusps carry the local degrees of freedom of the theory and correspond to the parabolic points
of a dense embedding in the functional integral of twisted parabolic Higgs bundles (of a non-commutative
version of the large-$N$ $YM$ theory).
From a stringy point of view the cusps are the $D$-branes at which open strings
can end \cite{Pen}.
This also explains a posteriori why, to get localization in the loop equation, a change of variables is needed
in which the functional integral is defined at the points rather than at 
the links of a lattice.
This means, in continuum language, that in our version of the loop equation we change variables from the connection 
to (the $ASD$ part of) the curvature. \par
Parabolic Higgs bundles have been introduced in the functional integral of pure $YM$ theory in  \cite{MB2}
and more recently in the functional integral of $\cal{N}$ $=4$ $SUSY$ $YM$ theory in \cite{W2} in their study
of the ramified Langlands conjecture.
We find that the exact beta function arises localizing the quasi $BPS$  Wilson loops 
on the moduli of twisted parabolic Higgs bundles of a special type
that correspond to the $Z_N$ non-abelian vortices of a partial non-commutative large-$N$
Eguchi-Kawai ($EK$) reduction \cite{EK,Neu,Twc,Twl1,Twl2}.  \par
On the physical side our result for the beta function
resembles the computation of the exact $\cal{N}$ $=1$
$SUSY$ $YM$ beta function from the evaluation of a very special sector
of the theory, the chiral sector, by the saddle-point method
via istantons \cite{NSVZ}. \par
We describe below some features of our approach from this physical point of view.
While summing all the large-$N$ diagrams of the pure Yang-Mills theory is a very difficult task,
perhaps outside the limits of our present techniques, restricting to diagrams
that contain only charge renormalization might result in a much simpler problem.
Indeed experience with supersymmetric gauge theories, even with only $\cal{N}$ $=1$
supersymmetry, suggests that the scheme dependence of the beta function
may be exploited in such a way to arrive at exact results, such as
the $NSVZ$ exact beta function of $\cal{N}$ $=1$ $SUSY$ Yang-Mills theory.
The original derivation of the $NSVZ$ exact beta function
relies crucially on instantons computations and on the 
cancellations between bosonic and fermionic non-zero modes
due to the $\cal{N}$ $=1$ supersymmetry \cite{NSVZ}. \par
However we can still look at the $NSVZ$ result as an ingenious way
to exploit the scheme dependence of the beta function in order to sum all the loop
diagrams that contribute to charge renormalization by means
of what is a one-loop computation for the
Wilsonean coupling and a (one-loop) rescaling anomaly for the canonical 
coupling of the $\cal{N}$ $=1$ $SUSY$ gauge theory \cite{AA}.
Yet in the pure $YM$ theory there is no supersymmetry. 
Thus our starting point, to simplify the problem, is the large-$N$ limit and
the loop equation. \par
Solving the loop equation uniformly for all Wilson loops is equivalent to find the master
field \cite{W3}, yet a too difficult problem \cite{MB1}, equivalent to summing all the large-$N$ diagrams.
Instead we want to restrict ourselves to charge renormalization only.
Thus we want to find a Wilson loop whose only renormalization
is charge renormalization in its internal loop diagrams.
We thus require that the loop is simple, i.e. without self-intersections,
and we also require that it has no perimeter divergence.
In addition we require that it has no cusp anomaly for backtracking cusps
since we want to use eventually the zig-zag symmetry (i.e. the invariance of the Wilson loop
by adding a backtracking arc) in the loop
equation. \par
Let us suppose that such a Wilson loop exists.
We choose a simple loop
and we insert it in the Migdal-Makeenko ($MM$) version of the loop equation \cite{MM,MM1}.
By loop equation we mean here the identity obtained requiring that the integral
of the functional derivative of the Wilson loop times the exponential of the action vanishes,
not its stringy version in terms of the loop operator \cite{MM,MM1}. \par
In the $MM$ loop equation, when the equation of motion is inserted in front of the 
Wilson loop in the left hand side,
a contact term is produced in the right hand side \cite{MM,MM1}.
This contact term is responsible of all quantum corrections in the large-$N$ limit,
in an iterative perturbative solution of the loop equation.
The contact term gives rise to perimeter and cusp divergences in the right hand side
of the loop equation \cite{Gr}.
Hence, because of its non-renormalization properties,
we would expect that the right hand side of the $MM$ loop equation
is zero when the quasi $BPS$ Wilson loop is inserted. \par
However this is not the case.
The $MM$ loop equation does not see any difference between the quasi $BPS$ Wilson loop
and an ordinary unitary Wilson loop.
The quantum contact term is anyway generated and cancellations for the quasi $BPS$ Wilson loop
occur only in the
iterative perturbative solution. \par
The situation gets even worse in presence of backtracking arcs ending into a cusp,
because the arcs, having a length, contribute to both the perimeter
divergence and the cusp anomaly in the right hand side of the loop equation.
The question arises as to how we can reconcile the cusp anomaly with
the zig-zag symmetry.
For an ordinary unitary Wilson loop
the way out is that the perimeter divergence in the right hand side of the loop
equation
has to be partially cancelled by an infinite 
cusp anomaly to recover the zig-zag symmetry,
in such a way that the only perimeter divergence that remains
is the one of the loop without the backtracking arcs.
However this cancellation is highly regularization (i.e. scheme)
dependent and essentially it is fixed assuming the zig-zag symmetry.
Indeed, despite the zig-zag symmetry of the loop,
the $MM$ loop equation is not regularized in a manifestly zig-zag invariant way,
since the perimeter and the cusp contribution in its
right hand side are separate.
Of course the same argument applies to the quasi $BPS$ Wilson loop,
but for the fact that in this case it should be possible to find a scheme 
in which the cusp anomaly cancels the perimeter divergence
completely. In fact it should be possible, possibly by a change of variables,
to find a regularization scheme in which the right hand side of the loop equation in the new variables
is actually zero
because of the mentioned non-renormalization properties. \par
We may expect that, if it exists a manifestly zig-zag invariant regularization of the new loop equation
(i.e. in a zig-zag invariant scheme),
its right hand side be zero for a quasi $BPS$ Wilson loop.
Indeed, since we already know that perimeter and cusp divergences have to mix together for backtracking
cusps, in a regularization in which there cannot be a cusp anomaly for backtracking cusps because of the manifest
zig-zag symmetry, there should not be a perimeter divergence either for the quasi $BPS$ Wilson loop.
Thus in such a scheme the right hand side of the loop equation is zero for a quasi $BPS$ Wilson loop.
But then, since charge renormalization occurs in the loop, it has to occur in the
left hand side of the loop equation. This means that it occurs in the effective action, that gives rise
to the equation of motion in the variables and regularization in which the new loop equation is written.
This effective action will contain all the quantum corrections for such a loop,
i.e. it will contain the sum of large-$N$ diagrams that renormalize the charge. \par
It turns out that a quasi $BPS$ Wilson loop exists in the large-$N$ limit of pure $YM$ theory
and that the variables that admit a zig-zag invariant regularization of the
loop equation are
the $ASD$ components of the curvature of the connection.
The zig-zag invariant regularization consists in analytically continuing the loop equation
from Euclidean to Minkowskian space-time.
It is perhaps at the heart of the modern Euclidean approach to quantum field theory
that such analytic continuation be possible. \par
In fact, taking advantage of the mixing between the perimeter and cusp divergences,
in our renormalization procedure every marked point of the loop is mapped into a backtracking cusp
by a singular conformal transformation that changes the renormalized action by the conformal
anomaly. \par
This is done by introducing a planar lattice, in a partial $EK$ non-commutative large-$N$
reduction of the theory from four to two dimensions, to which backtracking strings, that
do not change the loop by the zig-zag symmetry, are attached by means of a singular change of the conformal
structure \cite{Pen1}.
This suffices to define the beta function, from the divergences of the 
Euclidean effective action, and the regularized loop equation,
from the analytic continuation to Minkowskian space-time.
However, if we insist in keeping the $ASD$ structure real, the analytic continuation
of the Euclidean loop equation has actually to be performed to ultra-hyperbolic
signature. \par
It turns out that the loop equation is saturated by the Euclidean 
$ASD$ vortices of the non-commutative theory that, by consistency, admit analytic continuation to
ultra-hyperbolic signature
and that the (Euclidean) effective action computed on the vortices renormalizes with an exact beta 
function of $NSVZ$ type, whose first two coefficients agree with large-$N$ perturbation theory. \par
We should perhaps stress the twofold role played by attaching backtracking strings to the cusps,
i.e. to the parabolic points of our lattice.
At the level of loop equation the strings imply the absence of the contact term in the right hand
side of loop equation
because of the manifest zig-zag symmetry.
At the level of the functional integral the strings,
just because of their presence, allow the existence of a singular gauge
in which the (hermitean) curvature of the vortices of $ASD$ type can be diagonalized at each cusp.
In this singular gauge the original order of $N^2$ non-abelian integration variables 
at each cusp appear as order of $N^2$ zero modes of the Jacobian of the change of variables from
the connection to the $ASD$ curvature. Thus the Wilsonean beta function is
computed simply by counting zero modes as in the supersymmetric case, without actually performing
the order of $N^2$ integrations over the vortices moduli space:
\bea
\frac{1}{2 g^2_W(\tilde a)}
 =  \frac{1}{2 g^2_W(a)}- \frac{1}{(4\pi)^2} (2+\frac{5}{3})
log (\frac{\tilde a}{a})
\eea
where $a$ is a lattice cut-off and $\tilde a$ an infrared (lattice) scale.
However a crucial difference arises in the pure Yang-Mills case as
opposed to the $ \cal{N}$ $=1$ $SUSY$ case.
In the $SUSY$ case the whole Wilsonean beta
function is saturated by instantons zero modes.
In the pure $YM$ case only part of the beta function (the term equal to $2$) is accounted by the vortices zero modes.                                                                                                                                                                                  
The remaining part (the term equal to $\frac{5}{3}$) appears as a standard one-loop divergence 
of the Jacobian of the change of variables.
This divergence gives rise to a contribution of an anomalous dimension: 
\bea
\frac{\partial log Z}{\partial log \Lambda} =\frac{ \frac{1}{(4\pi)^2} \frac{10}{3} g_W^2}{1+cg_W^2}
\eea
in the canonical beta function, that is absent in the $SUSY$ case. \par
Indeed in the $ \cal{N} $ $=1$ $SUSY$ case the Jacobian of the change to the $ASD$ variables in the light-cone gauge
cancels (but for the zero modes) against the gluinos determinant.
In fact in the $ \cal{N} $ $=1$ $SUSY$ case and in the light-cone gauge our change of variables actually coincides with the Nicolai map
of the supersymmetric theory \cite{V, V1} \footnote{We would like to thank
Gabriele Veneziano for pointing out this feature to us.}.
The loop equation beta function is then saturated by instantons satisfying certain conditions,
thus reproducing the standard 
supersymmetric $NSVZ$ beta function in the large-$N$ limit \cite{MB3}. \par
The canonical beta function is found computing its relation to the Wilsonean one,
by rescaling the action into its canonical form. Indeed in the large-$N$ $YM$ 
theory there is a new contribution from the zero modes for this rescaling:
\bea
\frac{1}{2 g^2_W}
 = \frac{1}{2 g^2_c}+ \beta_J log g_c + \frac{\beta_J}{4} logZ
\eea
that upon differentiation by $\log(\frac{1}{a}) $ leads to the canonical beta function. \par
A posteriori there is a way of looking at the homological localization purely in terms of the 
flow of the $RG$.
In fact, remembering that the conformal map sends the marked points
to the cusps and that thus adds the conformal anomaly to the Wilsonean effective action,
the true renormalization of the Wisonean coupling at the cusps reads:
\bea
\frac{1}{2 g^2_W(\frac{ \tilde a}{\sqrt N_D})}
 =  \frac{1}{2 g^2_W(a)}- \frac{1}{(4\pi)^2} (2+\frac{5}{3})
log (\frac{\tilde a}{a\sqrt N_D})
\eea
where the large rescaling factor of $\sqrt N_D$ is due to the regularization of the singular
conformal anomaly at the cusps.
Thus the $ \log $ never gets large at the cusps, since the theory flows back to the ultraviolet
because of the conformal anomaly. The precise finite value of the  $ \log $ is just a matter of 
convention, i.e. a choice of the renormalization scheme at the cusps.\par
Thus at the cusps there is no infinite renormalization of the bare coupling.
Localization then occurs because at the ultraviolet the bare coupling vanishes.
Nevertheless, employing the $RG$ improved formulae for the anomalous dimension and the canonical beta function
obtained taking derivatives with respect to $\log(\frac{1}{a}) $, we 
get non-trivial formulae for the anomalous dimension and the canonical beta function. \par
This localization argument a posteriori
is based only on $AF$ and not on the particular value of the (first) coefficient
of the Wilsonean beta function. However this means that $AF$ is so difficult to produce
that if it occurs in the Wilsonean effective action obtained by a change of variables,
the resulting theory must necessarily be the $YM$ theory. \par
Coming back on the mathematical side, our result implies the existence, by explicit construction
in a certain regularization and renormalization scheme,
of the large-$N$ limit of the pure Yang Mills theory for the mentioned special quasi $BPS$ Wilson loops,
essentially because of the super-renormalizability of the large-$N$ Wilsonean
coupling, that turns out to be one-loop exact as in the $ \cal{N} $ $=1$ supersymmetric case 
and that implies an all orders 
formula of $NSVZ$ type for the canonical beta function \cite{MB1}. \par
We should perhaps specify in which sense we refer to the existence of the large-$N$ limit.
The existence of the functional integral can be shown either by abstract mathematical means
based on constructive quantum field theory or by direct construction of a solution. 
The constructive quantum field theory approach has a long tradition \cite{GA}.
Instead the techniques presented in this paper lead naturally
to the second point of view
together with the idea that the functional integral is defined by the way of computing
it, in this case by the solution of the large-$N$ loop equation. \par
We should perhaps mention that we construct a solution of the loop
equation for observables that, just because of their non-renormalization
properties, are physically almost trivial (at least in perturbation theory). It is precisely this almost triviality
that allows us to reach a solution, because most of the vast information
contained in the complete large-$N$ solution is lost.
Nevertheless we are able to extract from our solution the exact beta function,
that does not depend on a specific observable because of the universal character
of the renormalization procedure. \par
The plan of the paper is as follows. \par
In sect.2 we show the existence of the quasi $BPS$ Wilson loop
in large-$N$ pure $YM$ theory starting by analogy with the case of extended supersymmetry.
A few comments are in order. In the case of extended supersymmetry
$BPS$ Wilson loops are obtained adding to the gauge connection (scalar) Higgs
fields with a factor of $i$ in front \cite{Mal}, in order to satisfy non-trivially
the supersymmetric constraints. \par
Now already in $ \cal{N} $ $=1$ $d=4$ $SUSY$ $YM$ gauge theory there are no scalars
to play with.
However in such a theory we can construct a planar $BPS$ Wilson loop adding
to the gauge connection on the $(01)$ plane the covariant derivatives
along the orthogonal $(23)$ plane, that transform as an Higgs field for gauge transformations \cite{MB1}.
In fact this holds more properly
in a non-commutative version of the theory in which the derivative part of
the covariant derivative can be absorbed into a gauge transformation,
since in gauge theories on non-commutative space the gauge group contains the translations \cite{RT,Mak1,Mak2}. 
In turn this feature of absorbing translations into gauge transformations 
basically leads to the $EK$ large-$N$ reduction \cite{RT,Mak1,Mak2}. \par
Now, although there is no notion of supersymmetry, the very same operator
exists also in pure non-commutative $YM$ theory.
Thus while in theories with $ \cal{N} $ $=4,2$ $SUSY$ we obtain the Higgs field
needed for the $BPS$ property via dimensional reduction from the ten or the six dimensional $ \cal{N} $ $=1$ $SUSY$ theory
that is actually used to construct the extended supersymmetry in $d=4$,
in $ \cal{N} $ $=1,0$ $SUSY$ theories we obtain the Higgs fields via large-$N$
$EK$ reduction by means of covariant derivatives. \par
Now, as we show in sect.2, if we limit ourselves to the perimeter and the cusp divergences,
in the $ \cal{N} $ $=1$ $SUSY$ and in the pure $YM$ theory case, the mentioned
non-renormalization properties hold in the large-$N$ limit
not because of the supersymmetry but because of the $O(4)$ ($O(2,2)$ in ultra-hyperbolic signature)
symmetry of the non-commutative theory
in the limit of infinite non-commutativity. \par
In sect.3 we recall the loop equation of the pure $YM$ theory in the $ASD$ variables
following \cite{MB1} . We defer the study of the loop equation in the $ \cal{N} $ $=1$
case to another paper \cite{MB3}, where we interpret
our change of variables in the pure $YM$ theory as the Nicolai map in $ \cal{N} $ $=1$ $SUSY$ \footnote{We would like to thank
Gabriele Veneziano for pointing out this feature to us.}.
In fact it turns out that in the $ \cal{N} $ $=1$ case our loop equation leads to 
localization on the non-commutative (ultra-hyperbolic) instantons in the limit of infinite
non-commutativity and to the already known exact $NSVZ$ beta function in the large-$N$ limit \cite{MB3}. \par
In sect.4 we compute the beta function of the pure large-$N$ $YM$ theory
generalizing and simplifying in several ways
the computation already presented in \cite{MB1} . 
The basic fact is that the loop equation and the beta function are saturated by
the non-commutative  $ASD$ (ultra-hyperbolic) vortices of the $EK$ reduction in the pure $YM$ case as
opposed to the non- 
commutative (ultra-hyperbolic) instantons of the $ \cal{N} $ $=1$ case. \par
In sect.5 we write down rather explicitly our beta function and we exploit the residual
scheme dependence to create a link with a certain definition of the physical effective 
charge \cite{L} in the inter-quark potential in the large-$N$ limit.
We also compare our large-$N$ beta function with the numerical results found for $SU(3)$ \cite{L}. \par
In sect.6 we recall our conclusions.

\section{Quasi $BPS$ Wilson loops}

We recall in this section the properties of (locally) $BPS$ Wilson loops \cite{Mal}
originally introduced in the study of $ \cal{N} $ $=4$ $SUSY$ $YM/AdS$ string duality in \cite{Mal1}.
We use the notation and some of the arguments of \cite{Gr} 
about non-renormalization
properties of locally $BPS$ Wilson loops. \par
It has been argued in \cite{Gr} that a locally $BPS$ Wilson loop in the four-dimensional $ \cal{N} $ $=4$ $SUSY$ gauge
theory:
\bea
Tr\Psi(BPS)=Tr P \exp i\int_{C} (A_{a} dx_{a}(s)+i\phi_{b} dy_{b}(s))
\eea
has no perimeter divergence to all orders in perturbation theory because of the local $BPS$ constraint:
\bea
\sum_{a} \dot x^2_{a}(s)-\sum_{b} \dot y^2_{b}(s)=0
\eea
Indeed at lowest order of perturbation theory this constraint assures the cancellation
of the contribution to the perimeter divergence of the gauge propagator versus the scalar propagator, because of the 
factor of $i^2$ in front of the scalar propagator at that order.
As far as the perimeter divergence is concerned it is argued in \cite{Gr} that this cancellation occurs
to all orders in perturbation theory, when the locally $BPS$ Wilson loop is seen as the dimensional reduction
to four dimensions of the ten-dimensional Wilson loop of the ten-dimensional $ \cal{N} $ $=1$ $SUSY$ $YM$ theory
from which the four-dimensional $ \cal{N} $ $=4$ $SUSY$ theory is obtained. \par
We report here only the ten-dimensional version of the argument in \cite{Gr}, in which
$SUSY$ plays no role. In fact the argument is based only on $O(10)$ rotational symmetry,
as we show momentarily. In ten dimensions the coefficient of the perimeter divergence
of an ordinary unitary Wilson loop at any order in perturbation theory
must necessarily contain as a factor a polynomial in the $O(10)$ invariant 
quantity $\sum_{\alpha} \dot x^2_{\alpha}(s)$ without a constant term.
Indeed at any order in perturbation theory
a generic contribution to the Wilson loop contains a correlator of gauge fields,
i.e. a Green function, with tensor indices contracted with a product of monomials in
$\dot x_{\alpha}(s)$ at generic insertion points on the loop, labeled by $s$. \par
The perimeter divergence arises
when all insertion points coincide in such a way that all the arguments
of the Green function vanish. In this case the Green function provides
a factor that by $O(10)$ rotational invariance must be a polynomial
in ten-dimensional Kronecker delta, since all the difference vectors are zero at coinciding points
and thus no other tensorial structure can be produced. \par
This combines with the factors of $\dot x_{\alpha}(s)$ to produce an invariant polynomial
in $\sum_{\alpha} \dot x^2_{\alpha}(s)$ with no constant term since the lowest order
contribution is zero by direct computation.
But $\sum_{\alpha} \dot x^2_{\alpha}(s)$  is zero for a $BPS$ Wilson loop because of
the $BPS$ constraint.
A naive application of this argument to the four-dimensional $BPS$ Wilson loop
would imply the absence of the perimeter divergence for this loop on the basis of the $O(10)$
rotational invariance of the theory before the dimensional reduction.
Independently of subtleties eventually needed to apply the argument to 
the dimensionally reduced $d=4$ Wilson loop in the $ \cal{N} $ $=4$ theory we show now that
in $ \cal{N} $ $=1$ and $ \cal{N} $ $=0$ theories Wilson loops exist for which no perimeter
divergence occurs because of the $O(4)$ rotational symmetry in the large-$N$ limit.
We will show momentarily why in absence of $SUSY$ the large-$N$ limit is needed. \par
The connection that we look for is the following one:
\bea
B=A+D=(A_z+D_u) dz+(A_{\bar z}+ D_{\bar u}) d \bar z
\eea
where $z,\bar z,u,\bar u$ are the four-dimensional complex coordinates and $D_u=\partial_u
+i A_u$ the $u$ component of the covariant derivative.
From a two-dimensional point of view $B$ is a non-hermitean connection defined as the sum of a hermitean 
two-dimensional connection and a Higgs field. Indeed covariant derivatives transform as a Higgs field for gauge transformations.
Let us notice also that in the $ \cal{N} $ $=1$ theory $B$ gives rise to 
a locally $BPS$ Wilson loop because of the constraint that is implicit in its definition:
\bea
dz(s)=du(s) \nonumber \\
d\bar z(s)= d \bar u(s)
\eea
The factor of $i$ in front of the connection in the covariant derivative
plays the same role as the factor of $i$ in front of the Higgs field
in locally $BPS$ Wilson loops of theories with extended $SUSY$.
If the derivative term were absent in the covariant derivatives we could invoke the $O(4)$ symmetry
of the theory and the local $BPS$ constraint 
to imply the absence of the perimeter divergence as in the $O(10)$ $SUSY$ case.
Yet, there are derivatives. \par
In fact if we make the space non-commutative in the $u, \bar u$  directions,
we can get rid of the derivatives by a gauge transformation because the
translations can be absorbed into the gauge transformations
in a non-commutative theory \cite{DN} and more generally in the large-$N$ limit \cite{RT}. \par
However non-commutativity breaks the $O(4)$ symmetry in such a way that the property that we
are looking for is lost. \par
Yet, since the limit of infinite non-commutativity is equivalent to the ordinary
large-$N$ limit \cite{DN}, as can be seen for example from the loop equation of the non-commutative theory,
and in the ordinary commutative large-$N$ limit certainly $O(4)$
symmetry holds, then in the limit of infinite non-commutativity
our Wilson loop has no perimeter divergence. \par
This holds in the pure $YM$ theory and in the $ \cal{N} $ $=1$ $YM$ theory without using $SUSY$.
We now show, by an argument based on the $MM$ loop equation, that 
the perimeter divergence and the cusp anomaly for backtracking cusps have to mix
for a zig-zag invariant Wilson loop in such a way that the cusp anomaly cancels the
part of the perimeter divergence due to the backtracking arcs.
An argument of this kind has been anticipated in \cite{Gr}  in their discussion of the zig-zag symmetry 
on the stringy side of the correspondence $AdS/CFT$.\par
A byproduct of this argument is that a zig-zag invariant loop that has no perimeter 
divergence has no cusp divergence for backtracking cusps, because the zig-zag
symmetry requires cancellation between the two contributions and since one of them
vanishes the other one has to vanish too. \par 
Introducing the $MM$ loop equation at this stage will allow us to prepare our arguments on charge
renormalization too.
We can write the $MM$ loop equation for unitary Wilson loops in the large-$N$ $YM$ theory as:
\bea
\int_{C(x,x)} dx_{\alpha}\tau(\frac{1}{2 g^2}\frac{\delta S}{\delta A_{\alpha}(x)}\Psi(x,x;A))= \nonumber \\
i \int_{C(x,x)} dx_{\alpha} \int_{C(x,x)} dy_{\alpha} \delta^{(4)}(x-y) \tau(\Psi(x,y;A)) \tau(\Psi(y,x;A)) 
\eea
where the normalized trace, $\tau$, is the combination of the normalized v.e.v. with the normalized
colour trace in the fundamental representation (see for example \cite{MB1}) and
\bea
\Psi(x,y;A)=P \exp i\int_{C_{(x,y)}} A_{\alpha} dx_{\alpha}
\eea
In the case of loops without self-intersections but with cusps the $MM$ loop equation
reduces to:
\bea
\int_{C(x,x)} dx_{\alpha}\tau(\frac{1}{2 g^2}\frac{\delta S}{\delta A_{\alpha}(x)}\Psi(x,x;A))= \nonumber \\
i \int_{C(x,x)} dx_{\alpha} \int_{C(x,x)} dy_{\alpha} \delta^{(4)}(x-y) \tau(\Psi(x,x;A)) 
\eea
Performing the two contour integrations along the loop in the right hand side, we get:
\bea
\int_{C(x,x)} dx_{\alpha}\tau(\frac{1}{2 g^2}\frac{\delta S}{\delta A_{\alpha}(x)}\Psi(x,x;A))= \nonumber \\
i(L a^{-3}+\sum_{cusp} \frac{ \cos \Omega_{cusp}}{ \sin \Omega_{cusp}} a^{-2})\tau(\Psi(x,x;A)) 
\eea
where $L$ is the perimeter of the loop and $\Omega_{cusp}$ the cusp angle at a cusp.
The perimeter divergence arises by the double integration of the four-dimensional delta function,
i.e. of the contact term, along the loop.
However integrating the contact term in a neighborhood of each cusp
gives rise to a sub-leading quadratic divergence, since around a cusp
we get two independent integrations instead of one, due to the two
sides of the cusp. The coefficient of the cusp 
contribution is proportional to the ratio:
\bea
\frac{\cos \Omega_{cusp}}{\sin \Omega_{cusp}} 
\eea
The numerator arises from the scalar product,
the denominator from the two independent integrations of the two-dimensional
delta function.
In the limit in which the cusp angle $ \Omega_{cusp} $ reaches $\pi$, i.e. 
the cusp backtracks, the cusp contribution in the
contact term of the ordinary loop equation is negative and divergent. \par
If the Wilson loop is zig-zag invariant and the theory is regularized in a zig-zag invariant way 
this divergence must be fine-tuned to cancel
part of the perimeter divergence due to the length of the cusps, in such a way that the only
remaining perimeter divergence is the one of the loop without backtracking cusps, since these cusps
are irrelevant because of the zig-zag symmetry.
Thus we conclude that if a loop is zig-zag symmetric and it has no perimeter divergence it cannot have
a cusp anomaly either for backtracking cusps.
We can write the $MM$ loop equation also for a planar quasi $BPS$ Wilson loop:
\bea
\int_{C(x,x)} d\bar z\tau(\frac{1}{2 g^2}\frac{\delta S}{\delta B_z(x)}\Psi(x,x;B))= \nonumber \\
i\delta^{(2)}(0) \int_{C(x,x)} d\bar z\int_{C(x,x)} dz \delta^{(2)}(z-x) \tau(\Psi(x,z;B)) \tau(\Psi(z,x;B)))
\eea
For quasi $BPS$ Wilson loops the contact term in the
loop equation is the same as for unitary Wilson loops.
Cancellations occur only in the solution. \par
Hence the $MM$ loop equation cannot localize, not even for a quasi
$BPS$ Wilson loop, because it cannot be regularized in a way that does implement
the zig-zag symmetry explicitly.
In the next section we write a new loop equation for quasi $BPS$ Wilson loops
in which the zig-zag symmetry can be implemented explicitly.

\section{Localization by homology of the loop equation in the $ASD$ variables}

Cohomological localization in quantum field theory \cite{W1} is based on the fact that the integral of
an equivariantly closed differential form depends only on its cohomology class
and not on a particular representative:
\bea
\int (\omega+ Q \alpha)= \int \omega
\eea
Thus we have the freedom to add to the exponential of a closed form $\omega$, the action of our
quantum field theory, the $BRST$ differential of any globally
defined form, with an arbitrary coupling $t$, without changing the integral: 
\bea
\int \exp(\omega+ t Q \alpha)
\eea
as it is seen differentiating with respect to $t$ and using Eq.(22).
In this way the saddle-point
method applies exactly to the modified action in the limit $ t \rightarrow \infty $.
Localization of the functional integral on the critical points of $Q \alpha$ then follows. 
From this argument it is clear that localization is not a property of all the observables of the
theory but of only those that are equivariantly closed as the action is. \par
We would like to find an analog of cohomological localization 
for the pure $YM$ theory. Unfortunately $YM$ theory has no (twisted) $SUSY$ and thus it is unlikely
that this theory will ever admit a cohomological localization of any kind. \par
As a possible way out we look at the large-$N$ loop equation, the $MM$ equation.
In the $MM$ equation the contact term is the obstruction to localization,
since in its absence the $MM$ equation would reduce to the insertion in front of the Wilson loop
of the classical equation of motion of the theory, that is the saddle-point equation 
for the action.
Of course it is precisely this obstruction that makes the $MM$ equation
non-trivial and interesting.
By localization of the loop equation we mean a transformation of the loop equation
into the new form:
\bea
\tau(\frac{\delta \Gamma_q}{\delta  \mu(x)} \Psi(x,x;B)) =0
\eea
for an effective action, $\Gamma_q$, for a Wilson loop of a special kind, $B$,
and a trace, $\tau$, involving some new, yet unknown, integration variable, $\mu$.
Thus the transformation that we look for must eliminate the contact term,
after the change of variables from the gauge
connection to the yet unspecified field $\mu$. \par
We can think of the $MM$ equation as an equation defined on loops rather than on points.
Thus we will attempt to localize the loop equation by homological 
rather than cohomological methods. \par
Our basic homological property will be the zig-zag symmetry, i.e. the freedom to add to a Wilson loop
a backtracking arc without changing its holonomy \cite{Pol}.
This symmetry follows by the definition of the path-ordered exponential.
The zig-zag symmetry is the homological analog in our context of the cohomological identity in
Eq.(22). \par
For ordinary Wilson loops
the regularization and renormalization procedure may spoil the zig-zag symmetry.
In fact we have already seen that the perimeter and cusp
divergences of a smooth loop and of the same loop with the addition of the boundary of 
a tiny strip ending into a cusp may not coincide.
A limit case is dimensional regularization, in which there is no perimeter divergence,
since there are no linear divergences in $YM$ theory in this regularization,
but there is a logarithmic cusp divergence.
We saw however that we can still implement the zig-zag symmetry in an another regularization
by fine-tuning
the perimeter and cusp divergences in such a way that they partially cancel each
other in order to maintain the zig-zag symmetry.
In any case we saw that for the quasi $BPS$ Wilson loops introduced in the previous section
there is no perimeter divergence and no cusp anomaly for backtracking cusps.
Thus the basic homological identity holds for them without worrying about regularization
and renormalization. \par
In fact the zig-zag symmetry is by far too general to lead alone to homological localization.
The next (dynamical) ingredient, needed to get homological localization, is
a symmetry of the theory that generates the holonomically trivial deformation of a quasi
$BPS$ Wilson loop. This is analogous in cohomology to a symmetry of the action
that generates a deformation by a co-boundary, i.e. to the action being a closed form. 
As we will see below, in the classical $YM$ theory there is such a symmetry, it is the conformal symmetry.
However just because it is a symmetry of the classical theory, it will be of no use for our homological
localization.
What we need is a symmetry of the $RG$
flow of the (Wilsonean) renormalized effective action. This symmetry arises as follows.
There is an action of the two-dimensional conformal group that adds to the quasi $BPS$ Wilson loop
the boundary of a tiny strip \cite{Pen1}, that lifts to a four-dimensional conformal rescaling because of the diagonal
embedding of the planar quasi $BPS$ Wilson loop in the four-dimensional theory. \par	
At this point we can get localization in the loop equation simply by drawing a picture.
We must add to the Wilson loop a family of weighted arcs \cite{Pen} in such a way that all the marked points of the loop
are mapped into backtracking cusps \cite{Pen1}.
This modification amounts to a (singular) change of the conformal structure
around the marked points \cite{Pen1}. In the four-dimensional $YM$ theory this generates a conformal anomaly
in the effective action.
Since the cusps are backtracking their contribution in the right hand side of a new loop equation
in new variables will vanish, provided
the new loop equation can be regularized in a way compatible with the zig-zag symmetry.
Homological localization follows. \par 
The family of arcs is not arbitrary, since it must be compatible with the gluing properties
of the functional integral. This leads to the axioms of the arc complex of topological closed/open
strings \cite{Pen} as we will see below. \par
The argument that we have presented hides a subtle but crucial point that will lead us to the right change of variables.
We can choose the marked point of a loop
arbitrarily. This implies that we would need uncountably many arcs and cusps.
Zig-zag symmetry applies instead to a family finitely generated.
To get a finitely generated family of arcs we can introduce a finite lattice of cusps and then take the continuum limit
as in lattice gauge 
theories. But in lattice gauge theories the integration variables live on the links of a lattice rather than on the
points. This is the crucial point. We must change variables from the connection to the curvature
for homological localization to apply.
The curvature lives on the plaquettes, but they are dual to the points in two dimensions.
Thus we need a partial $EK$ reduction from four to two dimensions. \par
In addition the curvature has too many components for a change of variables to exist in four dimensions,
since the gauge connection has four components and the curvature six.
The $ASD$ part of the curvature has only three components.
Thus we need the choice of an axial gauge to kill one component of the connection.
In four dimensions there are reason to prefer the light-cone gauge
\cite{Bas}.
In the light-cone gauge the number of components matches. In fact this change of variables in the $SUSY$ $YM$
theory is the Nicolai map \footnote{We thank Gabriele Veneziano for pointing out this feature to us.}.
But the light-cone gauge exists only in Minkoskian space-time.
This suggests that the localization of the loop equation of quasi $BPS$ Wilson loops occurs only 
if the loop equation, that is gauge invariant, is written in the $ASD$ variables in a way that admits analytic continuation
to Minkowskian space-time and if the theory is regularized on a lattice. \par
We now write the formulae that correspond to our arguments.
We can describe the chain of changes of variables
and transformations that lead to large-$N$ homological localization as follows. 
We start with the large-$N$ $YM$ theory defined on  $R^2 \times R_{\theta}^2$  in the limit of infinite non-commutativity
$\theta$, that is known to reproduce the ordinary commutative large-$N$ limit:
\bea
Z=\int \exp(-\frac{N}{2g^2} \sum_{\alpha \neq \beta} \int Tr_f (F_{\alpha \beta}^2) d^4x) DA \nonumber \\
=\int \exp(-\frac{N 8 \pi^2 }{g^2} Q-\frac{N}{4g^2} \sum_{\alpha \neq \beta} \int Tr_f(F^{-2}_{\alpha \beta}) d^4x) DA
\eea
In the second line the classical action is conveniently rewritten as the sum of a topological and a purely $ASD$ term.
The topological term $Q$ is the second Chern class, given by:
\bea
Q=\frac{1}{16 \pi^2} \sum_{\alpha \neq \beta} \int Tr_f (F_{\alpha \beta} \tilde F_{\alpha \beta}) d^4x
\eea
while the $ASD$ curvature $F^-_{\alpha \beta}$ is defined by:
\bea
F^-_{\alpha \beta}=F_{\alpha \beta}- \tilde F_{\alpha \beta} \nonumber \\
\tilde F_{\alpha \beta} = \frac{1}{2} \epsilon_{\alpha \beta \gamma \delta} F_{\alpha \beta}
\eea
Introducing the projectors, $P^-$ and $P^+$, the curvature can be decomposed into its
$ASD$ and $SD$  components:
\bea
F_{\alpha \beta}=P^-F_{\alpha \beta}+P^+ F_{\alpha \beta} \nonumber \\
= \frac{1}{2}F^-_{\alpha \beta}+\frac{1}{2}F^+_{\alpha \beta}
\eea
Notice that the coefficient of the classical action as a functional of the projected $ASD$ curvature is
twice the coefficient of the classical action as a functional of the total curvature.
This will be important when we will compute the beta function.
The generators in the fundamental representation are normalized as:
\bea
Tr_{f}(T^a T^b)= \frac{1}{2} \delta_{ab} \nonumber \\
\sum_a (T^a)_{f}^2=\frac{N^2-1}{2 N} 1_{f}
\eea
We change variables from the connection to the $ASD$ curvature,
introducing in the functional integral the appropriate resolution of identity:
\bea
1= \int \delta(F^{-}_{\alpha \beta}-\mu^{-}_{\alpha \beta}) D\mu^{-}_{\alpha \beta}
\eea
The partition function thus becomes:
\bea
Z=\int \exp(-\frac{N 8 \pi^2 }{g^2} Q-\frac{N}{4g^2} \sum_{\alpha \neq \beta} \int Tr(\mu^{-2}_{\alpha \beta}) d^4x)
\nonumber \\
\times \delta(F^{-}_{\alpha \beta}-\mu^{-}_{\alpha \beta}) D\mu^{-}_{\alpha \beta} DA
\eea
We can write the partition function in the new form:
\bea
Z=\int \exp(-\frac{N 8 \pi^2 }{g^2} Q-\frac{N}{4g^2} \sum_{\alpha \neq \beta} \int Tr(\mu^{-2}_{\alpha \beta}) d^4x)
\nonumber \\
\times Det'^{-\frac{1}{2}}(-\Delta_A \delta_{\alpha \beta} + D_{\alpha} D_{\beta} +i ad_{\mu^-_{\alpha \beta}} )
D\mu^{-}_{\alpha \beta}
\eea
where the integral over the gauge connection of the delta function has been now explicitly performed:
\bea
\int DA_{\alpha} \delta(F^-_{\alpha \beta}- \mu ^-_{\alpha \beta})= |Det'^{-1}(P^- d_A \wedge)|
\nonumber \\
=Det'^{-\frac{1}{2}} ((P^- d_A \wedge)^*(P^- d_A \wedge)) \nonumber \\
=Det'^{-\frac{1}{2}}(-\Delta_A \delta_{\alpha \beta} + D_{\alpha} D_{\beta} +i ad_{ F^-_{\alpha \beta}} )
\eea
and, by an abuse of notation,
the connection $A$ in the determinants denotes the solution of the equation
$F^-_{\alpha \beta}- \mu ^-_{\alpha \beta}=0$.
The $ ' $ superscript requires projecting away from the determinants the zero modes due to gauge invariance, since
gauge fixing is not yet implied, though it may be understood
if we like to. \par
We refer to the determinant in the preceding equation as to the localization determinant,
because it arises localizing the
gauge connection on a given level, $\mu ^-_{\alpha \beta}$, of the $ASD$ curvature.
Let us notice the unusual spin term $i ad_{ F^-_{\alpha \beta}}$ as opposed to the one that arises
in the background field method $2i ad_{ F_{\alpha \beta}}$. 
In the background field method the quadratic form:
\bea
\frac{1}{2g^2}\int d^4x Tr(F_{\alpha \beta})^2
\eea
is expanded around a solution of the equation of motion.
This produces the factor of two in the spin term.
The localization determinant arises instead from the expansion of the quadratic form:
\bea
\lim_{\epsilon \rightarrow 0} \frac{1}{\epsilon} \int d^4x Tr(F^{-}_{\alpha \beta}-\mu^{-}_{\alpha \beta})^2
\eea
that arises in a definition of the delta function around the background:
\bea
F^{-}_{\alpha \beta}=\mu^{-}_{\alpha \beta}
\eea
The corresponding shift in the curvature
explains why the coefficient of the spin term is one half
of the one in the background field method. The occurrence
of $F^{-}_{\alpha \beta}$ is due instead to the projector on the $ASD$ part of the curvature. \par
After this change of variables the functional integral is defined on the plaquettes rather than on the
links of a lattice, in a lattice regularization of the theory that will be needed later. \par
We need another change of variables to a holomorphic gauge, in order to write the loop equation in 
the $ASD$ variables in a way convenient to our further developments.
This introduces a holomorphic anomaly in the functional integral as a Jacobian.
The choice of a holomorphic gauge is required for the following reason.
We want to reduce the $MM$ loop equation, that is obviously defined on loops,
to a critical equation defined on points. There is a canonical way to associate
to a loop a point, via the evaluation of a residue.
The change of variables to the holomorphic gauge is meant to produce the
Cauchy kernel in the loop equation, that in turn can be evaluated as a 
regularized residue in a proper regularization. 
In fact the very idea of localization in the loop equation has a holographic interpretation \cite{MB1}
in which to a loop (with a marked point), on which the one dimensional quantum theory on the
bulk lives, it is associated a critical equation at the marked point, on which the zero dimensional localized
theory on the boundary lives. \par
We describe the holomorphic gauge as follows.
We can interpret the $ASD$ relations:
\bea
F^-_{\alpha \beta}- \mu ^-_{\alpha \beta}=0
\eea
as an equation for the curvature of the
non-Hermitean connection $B=A+D=(A_z+D_u) dz+(A_{\bar z}+ D_{\bar u}) d \bar z$ and a harmonic constraint for the Higgs field
$\Psi=-iD=-i( D_u dz+D_{\bar u} d \bar z )= \psi+\bar \psi$:
\bea
F_B - \mu=0 \nonumber \\
\bar F_{B} - \bar \mu =0 \nonumber \\
d^*_A \Psi - \nu =0
\eea
that can also be written as:
\bea
F_B - \mu=0 \nonumber \\
\bar \partial_A \psi- n=0 \nonumber \\
\partial_{A} \bar \psi- \bar n=0
\eea
where the fields $\mu, \nu, n$  are suitable linear combinations
of the $ASD$ components $\mu ^-_{\alpha \beta}$.
The resolution of identity in the functional integral
then reads:
\bea
1=\int \delta(F_B - \mu) \delta(\bar \partial_A \psi- n)  \delta(\partial_{A} \bar \psi- \bar n)
D \mu Dn D \bar n
\eea
where the measure $D \mu$ is interpreted in the sense of holomorphic matrix models \cite{Laz},
employed in the study of the chiral ring of $ \cal {N}$ $=1$ $SUSY$ gauge theories \cite{Vafa}. \par
This interpretation of the measure $D \mu$ seems to be needed to get the correct counting, in order to reproduce the
perurbative beta function,
of (complex) zero modes in the effective action.
The holomorphic gauge is defined as the change of variables for the connection
$B$, in which the curvature of $B$ is given by the field $\mu'$,
obtained from the equation:
\bea
F_B - \mu=0
\eea
by means of a complexified gauge transformation $G(x;B)$ that puts
$B=b+ \bar b$ in the gauge $\bar b=0$:
\bea
\bar{\partial}b_z= -i \mu'
\eea
where $\mu'=G \mu G^{-1}$ (the factor of $i$ occurs because Eq.(42) is written in complex coordinates). \par
Employing Eq.(40) as a resolution of identity in the functional integral,
the partition function  becomes:
\bea
Z=\int
 \delta(F_B - \mu) \delta(\bar \partial_A \psi- n)  \delta(\partial_{A} \bar \psi- \bar n)
 \exp(-\frac{N}{2g^2}S_{YM}) \nonumber \\
 \times \frac{D \mu}{D \mu'}
Db D \bar b  D \mu' Dn D \bar n
\eea
The integral over $b, \bar b$ is the same as the integral over the four $A_{\alpha}$. The resulting functional
determinants, together with the Jacobian of the change of variables to the
holomorphic gauge, are absorbed into the definition of $\Gamma$. \par
$\Gamma$ plays here the role of a classical action, since it must be
still integrated over the fields $\mu', n, \bar n$.
We  call $\Gamma$ the classical
$ASD$  action, as opposed to the quantum $ASD$ effective action, $\Gamma_q$.
$\Gamma$ is given by:
\bea
\Gamma=\frac{N 8 \pi^2 }{g^2} Q
+\frac{N}{g^2} \int Tr_f(F^{-2}_{01}+F^{-2}_{02}+F^{-2}_{03} ) d^4x \nonumber \\
+ log Det'^{-\frac{1}{2}}(-\Delta_A \delta_{\alpha \beta} + D_{\alpha} D_{\beta} +i ad_{\mu^-_{\alpha \beta}})
-log\frac{D \mu}{D \mu'}
\eea
with:
\bea
\mu^0=F^-_{01}   \nonumber \\
n+\bar n=F^-_{02} \nonumber \\
i(n-\bar n)=F^-_{03}
\eea
Although $\Gamma$ is the classical action in the $ASD$ variables it contains already quantum 
corrections because of the Jacobian of the change of variables. It turns out that its divergent part
coincides with the divergent part of the Wilsonean
localized quantum effective action, after the inclusion of zero modes.
Until now the theory is still four dimensional.
Taking functional derivatives with respect to the $ASD$ field we get, for a planar quasi $BPS$ loop,
the loop equation:
\bea
0=\int  D\mu'  Tr \frac{\delta}{\delta \mu'(w,0)}
  (\exp(- \Gamma)
  \Psi(x,x;b)) \nonumber \\
 = \int  D\mu' \exp(-\Gamma)
  (Tr(\frac{\delta \Gamma}{\delta \mu'(w,0)} \Psi(x,x;b))
  \nonumber \\
  -\int_{C(x,x)} dy_z \frac{1}{2}  \delta^{(2)}(0) \bar{\partial}^{-1}(w-y)
  Tr(\lambda^a \Psi(x,y;b)
  \lambda^a \Psi(y,x;b)) ) \nonumber \\
 =\int D\mu' \exp(-\Gamma)
  (Tr(\frac{\delta \Gamma}{\delta \mu'(w,0)} \Psi(x,x;b)) \nonumber \\
  - \int_{C(x,x)} dy_z \frac{1}{2} \delta^{(2)}(0) \bar{\partial}^{-1}(w-y)(Tr( \Psi(x,y;b))
    Tr(\Psi(y,x;b))  \nonumber \\
  - \frac{1}{N} Tr( \Psi(x,y;b) \Psi(y,x;b))))
\eea
where in our notation we have omitted the integrations $D n D \bar n$ since they are irrelevant
in the loop equation because the curvature of $B$ depends only on $\mu$.
In the large-$N$ limit it reduces to:
\bea
 \tau(\frac{\delta \Gamma}{\delta \mu'(w.0)} \Psi(x,x;b)) = \nonumber \\
  \int_{C(x,x)} dy_z \frac{1}{2} \delta^{(2)}(0) \bar{\partial}^{-1}(w-y) \tau( \Psi(x,y;b))
    \tau(\Psi(y,x;b)))
\eea
The quadratically divergent factor $\delta^{(2)}(0)$ arises as follows.
The functional derivative at a point produces a factor of $\delta^{(4)}=\delta^{(2)} \times \delta^{(2)} $ 
at that point because the fields are four dimensional. One factor of $\delta^{(2)}$  is convoluted with the kernel of 
$\bar{\partial}^{-1}$, while the other factor produces $\delta^{(2)}(0)$ since the loop is assumed to
lie on a plane. \par
Now it is natural to perform a partial $EK$ reduction from four to two dimensions.
Let us describe what in fact the partial $EK$ reduction means in this context.
We already observed that we can absorb the translations into a gauge transformation
in the four-dimensional non-commutative theory along the two non-commutative directions \cite{DN},
that are the one transverse to the plane over which the loop lies.
As a result the classical action looks two dimensional, in the sense that the space-time dependence
of the fields
is two dimensional, despite the fact that the theory is truly four dimensional.
The four-dimensional information is hidden in the central extension  $H=\frac{1}{\theta}$
that shows up in the curvature of the connection because of the non-commutativity of the partial derivatives
in the non-commutative directions \cite{DN}.
However the loop equation, being gauge invariant, is unchanged by this special gauge choice
if the gauge is chosen after taking the functional derivatives.
This means that the quadratically divergent factor $\delta^{(2)}(0)$ occurs in the four-dimensional loop
equation even in a gauge in which the fields are constant in the non-commutative directions. \par 
If however we were to write the loop equation for the theory already reduced to two dimensions
we would miss the  $\delta^{(2)}(0)$  divergent factor in the right hand side, because we would get just 
one factor of $\delta^{(2)}$ by taking 
functional derivatives, instead of the factor of $\delta^{(4)}$. \par
Now we use the fact that in the non-commutative theory the integral over the non-commutative directions can be represented
as a (colour) trace \cite{DN}:
\bea
\int d^2x_T=\frac{2\pi}{H} Tr
\eea
Hence the non-commutative classical action of the reduced theory gets a volume factor of
$V_2=2\pi \theta=\frac{2\pi}{H}$ because of the gauge choice.
The equation of motion of the reduced theory is therefore multiplied by this volume factor.
We can divide both sides of the loop equation by this volume factor in the reduced theory
in such a way that the equation of motion is normalized as in the four-dimensional theory.
Then the inverse volume will appear in the right hand side instead of the factor $\delta^{(2)}(0)$ .
We can compensate this fact by rescaling the classical action by a factor of $N_2^{-1}$ \cite{RT},
with $N_2=V_2 \delta^{(2)}(0)$, in such a way that the factor of $\frac{V_2}{N_2}$ in the reduced classical action
produces the factor of $\delta^{(2)}(0)=
\frac{N_2}{V_2}$ once carried to the right hand side of the loop equation. 
Of course in all this discussion we are implicitly assuming that the trace of the reduced theory
includes now the non-commutative degrees of freedom.\par
Thus the reduced classical action of the twisted $EK$ reduced theory in the $ASD$ variables
is given by:
\bea
\Gamma=\frac{N 8 \pi^2 }{N_2 g^2} Q
+\frac{N}{g^2} \frac{2\pi}{N_2 H}\int Tr_f(F^{-2}_{01}+F^{-2}_{02}+F^{-2}_{03} ) d^2x \nonumber \\
+ log Det'^{-\frac{1}{2}}(-\Delta_A \delta_{\alpha \beta} + D_{\alpha} D_{\beta} +i ad_{\mu^-_{\alpha \beta}})
-log\frac{D \mu}{D \mu'}
\eea
where the trace in the functional determinants has to be interpreted
coherently with the partial $EK$ reduction. \par
The reduced loop equation is then:
\bea
\tau(\frac{\delta \Gamma}{\delta \mu'(w)} \Psi(x,x;b)) = \nonumber \\
   \int_{C(x,x)} dy_z \frac{1}{2} \bar{\partial}^{-1}(w-y) \tau( \Psi(x,y;b))
   \tau(\Psi(y,x;b)))
\eea
It is interesting to observe that the reducing $EK$ factor cancels exactly the quadratic divergence
obtained evaluating the action on parabolic Higgs bundles, once it is assumed that the transverse
and longitudinal cut-off are equal. But this is necessary to keep our $O(4)$ symmetry.
The partial $EK$ reduction is not strictly needed for our arguments and for computing
the beta function,
but it is a convenient technical tool. It allows us to avoid overall infinite factors to
appear in our formulae and by this very reason explains how apparently singular objects
as the parabolic Higgs bundles survive in the large-$N$ limit of the functional integral.
In this respect we should mention a different point of view between our treatment of 
parabolic Higgs bundles and the one in  \cite{W2}.
In fact in the mathematical literature we can think of parabolic bundles in two slightly different ways.
We can think that they are defined on a compact surface with a divisor and a parabolic structure that belongs
to the surface. This is our point of view.
Or we can think that they arise as boundary conditions on a surface with boundary.
This is the point of view in \cite{W2}. \par
After the $EK$ reduction from four to two dimensions we perform a conformal compactification
in such a way that the theory now is defined over a two-dimensional sphere. 
In the four-dimensional Euclidean theory this amounts to a compactification
from $R^4$ to $S^2 \times S^2$. In the four-dimensional theory in ultra-hyperbolic
signature the conformal compactification is instead from the Minkowski space-time $M$ to $\frac{S^2 \times S^2}{Z_2}$
\cite{Mas}.
This adds to the local part of the effective action at most a finite conformal anomaly, that can be ignored. \par
It is clear that the contour integration in the quantum term of the loop equation
includes the pole of the Cauchy
kernel. We need therefore a gauge invariant regularization.
The natural choice consists in analytically continuing the loop equation
from Euclidean to Minkowskian space-time. Thus $z \rightarrow i(x_+ + i \epsilon)$.
It is at the heart of the modern Euclidean approach to quantum field theory that
this analytic continuation be in fact possible.
This regularization has the great virtue of being manifestly gauge invariant.
In addition this regularization is not loop dependent. \par
The result of the $i \epsilon$ regularization of the Cauchy kernel is the sum of
two distributions, the principal part plus
a one-dimensional delta function (for simplicity we omit the underscript of $x_{+}$ in the following):
\bea
\frac{1}{2}\bar{\partial}^{-1}(w_x -y_x +i\epsilon)= (2 \pi)^{-1} (P(w_x -y_x)^{-1}
- i \pi \delta(w_x -y_x))
\eea
The loop equation thus regularized looks like:
\bea
\tau(\frac{\delta \Gamma}{\delta \mu'(w)} \Psi(x,x;b))= \nonumber \\
\int_{C(x,x)} dy_x(2 \pi )^{-1} (P(w_x -y_x)^{-1}
 - i \pi \delta(w_x -y_x)) \nonumber \\
\times \tau( \Psi(x,y;b)) \tau(\Psi(y,x;b)))
\eea
The right hand side of the loop equation
contains now two contributions.
A delta-like one dimensional contact term, that is supported on closed
loops and a principal part distribution that is supported 
on open loops. Since by gauge invariance it is consistent to assume
that the expectation value of open loops vanishes, the principal part 
does not contribute and the loop equation reduces to:  
\bea
\tau(\frac{\delta \Gamma}{\delta \mu'(w)} \Psi(x,x;b))= \nonumber \\
\int_{C(x,x)} dy_{x}
\frac {i}{2}\delta(w_x -y_x) \tau(\Psi(x,y;b)) \tau(\Psi(y,x;b)))
\eea
Taking $w=x$ and using the transformation properties of the holonomy of $b$
and of $\mu(x)'$, the preceding equation can be rewritten in terms
of the connection, $B$, and the curvature, $\mu$:
\bea
\tau(\frac{\delta \Gamma}{\delta \mu(x)} \Psi(x,x;B))= \nonumber \\
\int_{C(x,x)} dy_{x}
\frac {i}{2}\delta(x_x -y_x) \tau( \Psi(x,y;B)) \tau(\Psi(y,x;B)))
\eea
where we have used the condition that the trace of open loops vanishes
to substitute the $b$ holonomy with the $B$ holonomy. \par
Our argument about the vanishing of the contribution of the principal 
part becomes tricky in the non-exactly gauge invariant regularization
of the loop equation that was considered in \cite{Gr}.
Indeed that regularization allows the contribution of quasi-closed loops,
for which of course the principal part contributes too. 
Yet we will see that our final argument about reducing the loop equation to a saddle-point
on the weighted graph is not affected, not even if the contribution of the
principal part is allowed using the regularization in \cite{Gr}.
Indeed the zig-zag symmetry works for both the contact term and the principal
part. \par
We need a lattice version of the continuum loop equation
to implement our localization argument.
Thus we write the loop equation in the $ASD$ variables on a lattice in the partially $EK$ reduced theory.
On a dense set in the functional integral (in the sense of distributions), the equations:
\bea
F_{z \bar z}- [D_u,D_{\bar u}]=i (\sum_{p} \mu^0_p \delta^{(2)}(x-x_p)-H1)\nonumber \\
\bar{\partial}_A (D_u)=i \sum_{p}  n_p \delta^{(2)}(x-x_p)\nonumber \\
\partial_{A}(D_{\bar u})=i \sum_{p}  \bar{n}_p \delta^{(2)}(x-x_p)
\eea
define an infinite-dimensional
twisted local system or, what is the same, a twisted parabolic Higgs bundle on a sphere.
The curvature equation involves a central term, $H=\frac{1}{\theta}$,
that we can display explicitly since $[D_u,D_{\bar u}]= F_{u \bar u}+i H 1$.
Correspondingly, given that the gauge connection has to vanish at infinity also the
right hand side has been shifted by $H1$ by deforming the resolution of identity 
on the $ASD$ part of the curvature in the functional integral, as follows by  non-commutativity.
$H $ vanishes in the large-$N$ limit, that coincides with the limit of infinite
non-commutativity.
In the case $n=\bar n=0$, that will be the most relevant for us, we may interpret the preceding
equations as vortices equations.
In the partially $EK$ reduced theory the vortices live at the lattice points where the $ASD$ curvature
is singular. This means that in the original four-dimensional theory they form two-dimensional
vortices sheets \cite{MB2,W2}. 
The  loop equation on our lattice now reads:
\bea
\tau(\frac{\delta \Gamma}{ \delta \mu(x_p)}\Psi(x_p,x_p;B))=  \nonumber \\
\int_{C(x_p,x_p)} dy_z \frac{1}{2}\bar{\partial}^{-1}(x_p-y) \tau(\Psi(x_p,y;B))
\tau (\Psi(y,x_p;B)))
\eea
and correspondingly for the analytic continuation to Minkowskian space-time:
\bea
\tau(\frac{\delta \Gamma}{\delta \mu'(w_{x_q})} \Psi(x_q,x_q;B))=  \nonumber \\
 \int_{C(x_q,x_q)} dy_x(2 \pi )^{-1} (P(w_{x_q} -y_x)^{-1} 
  - i \pi \delta(w_{x_q} -y_x))  \nonumber \\
 \times  \tau( \Psi(x_q,y;B)) \tau(\Psi(y,x_q;B)))
\eea
Notice that a smooth marked point gives a non-trivial contribution to the right hand side of the loop equation in the
$ASD$ variables either in the continuum or on the lattice,
that is finite for the contact term and logarithmically divergent for the principal part (in the regularization in \cite{Gr}).
However around a backtracking cusp the contributions of the two sides of the asymptotes to the cusp
cancel each other for the contact term:
\bea
\int_{C(x_q,x_q)} dy_x(s)
\delta(w_{x_q}(s_{cusp}) -y_x(s))=\frac{1}{2} (\frac{\dot w_{x_q}(s^+_{cusp})}{ |\dot w_{x_q}(s^+_{cusp})|}+
\frac{\dot w_{x_q}(s^-_{cusp})}{|\dot w_{x_q}(s^-_{cusp})|}) 
\eea
because of the opposite sign of $ \dot w_{x_q}(s^+_{cusp})$ and $ \dot w_{x_q}(s^-_{cusp})$ 
on the two sides of the backtracking cusp.
For the principal part the same argument applies 
because of the opposite orientations of the asymptotes and because 
both the cusp asymptotes are approached either from below or from above:
\bea
|\int_{C(x_q,x_q)} dy_x(s)
P(w_{x_q}(s_{cusp}) -y_x(s))^{-1}|= \nonumber \\
\frac{1}{2} |\int ds^+ \frac{\dot y_x(s^+)}{|w_{x_q}(s_{cusp}) -y_x(s^+)|}+
\int ds^- \frac{ \dot y_x(s^-)}{|w_{x_q}(s_{cusp}) -y_x(s^-)|}| 
\eea
Thus if every marked point of the loop can be transformed into a backtracking cusp we can complete our argument
about localization.
But this is precisely the effect of our lattice, since marked points of the loop contribute to
the loop equation in the lattice theory only if they coincide with the lattice points.
Thus we can simply draw our backtracking strings from the loop to the lattice points
in order to transform all the marked points into cusps. \par
Hence we may say that open strings solve 
the $YM$ loop equation for the quasi $BPS$ Wilson loops, in the sense that they localize the 
loop equation on a saddle point for an effective action. \par 
We now show that the cusps must be paired by the backtracking strings. 
Independently on the string gluing axioms we can understand directly
from the gluing properties of the functional integral why it must be so. \par
We can represent the partition function on a sphere by gluing
the two cups with the Wilson loop and the marked point on the loop in common.
To apply our vanishing result for the contact term to the loop equation on each cup
we must connect the same marked point 
on each cup with a backtracking string ending into a cusp. If we now glue the two cups, 
the cusps come in pairs on the glued surface and are linked by a string transverse to the loop
that intersects the loop at the marked point.
These are precisely the string gluing axioms for arc families at topological level \cite{Pen}.
Thus we come to the conclusion that if we combine the gluing properties
of the functional integral with the vanishing requirement for the contact term, i.e.
the requirement of localization, we get the string gluing axioms at topological level.
A subtle point arises as follows.
When we add a backtracking arc to a marked point we create a loop self-intersection
that may contribute extra terms to the loop equation.
However the self-intersection may be regularized adding instead tiny strips
having a common asymptote to the two cusps. In this case the loop remains simple
and the cusps are still pairwise identified \cite{Pen}. 
Thus the proper graph to get localization is a weighted graph \cite{Pen} whose spine
is a Mandelstam graph. The weights in our language are the sizes of the strips. \par 
One more comment is in order.
It is one of the cornerstones of the Euclidean field theory that the analytic continuation to Minkowskian 
space-time can always be performed.
In fact if we want to keep the $ASD$ structure real we must continue to
ultra-hyperbolic signature. \par
If we perform the compactification to $S^2 \times S^2$ in Euclidean signature 
we get the double cover of the conformal compactification in ultra-hyperbolic signature \cite{Mas}.
Thus if we require that the Euclidean equation of $ASD$ type be continued to ultra-hyperbolic
signature we must take into account this global constraint. \par
We are now ready to write the quantum effective action for our localized version of the loop equation.
A subtle point arises about the cut-off of this effective action.
We recall that the introduction of a lattice is essential for localization, since it allows us to transform
every non-trivial marked lattice point into a backtracking cusp.
Now, because all lattice points are pairwise linked by strings and the strings are transverse to the loop,
there are the same number of lattice points 
inside and outside the loop. \par
Hence we need different cut-off scales to fix the two different areas of the cups in which the loop divides the sphere.
Thus the topological axioms imply different cut-offs
at the cusps of the two cups of the sphere in the quantum effective action. \par
In fact an equal cut-off would lead to overcounting in the normalization
of the classical action, that has to be the same as the one for a unique marked point,
up to terms vanishing with the ultraviolet cut-off. 
Hence the local part of the effective action, as a consequence of the stringy nature of localization in the loop
equation, is in fact bi-local with two local fields
(differing in fact only by a gauge transformation not connected to the identity
since they are originally associated to the same marked point by gluing)
living at different scales, one at the ultraviolet cut-off and one at the infrared cut-off.
The field at the ultraviolet, but not the one at the infrared,
affects the renormalization of the Wilsonean coupling constant,
as it is expected from its very definition.
Instead the field at the infrared together with the one at the ultraviolet affects the
renormalization of the canonical coupling constant. \par
Finally we draw our weighted graph, that is made by two charts with the boundary loop
in common. The charts are a conformal transformation of two topological disks with marked points.
This introduces a conformal transformation in the $EK$ two-dimensional reduced theory \cite{Pen,Pen1,Man}.
However this transformation is four dimensional in the original theory
because of the $BPS$ constraint extended to a neighborhood of the marked points:
\bea
dz=du \nonumber \\
d\bar z= d \bar u
\eea
Thus the four-dimensional metric changes conformally and
the effective action changes by the appropriate conformal
anomaly. This means that, in addition to the explicit cut-off dependence,
the effective action on the weighted graph is related to the one
on the sphere with marked points by the addition of a divergent conformal anomaly,
because of the singularity of the conformal transformation.
This divergent conformal anomaly plays a key role in our computation
of the contribution of the anomalous dimension in the canonical beta function
and more generally in the interpretation a posteriori of localization
as a $RG$ flow to the ultraviolet. \par 
The loop equation on the weighted graph reduces to the localized form:
\bea
0=\tau(\frac{\delta \Gamma_{q}}{\delta \mu(x_p)}\Psi(x_p,x_p;B))
\eea
which we refer to as the master equation.

\section{Effective action and exact beta function}

In this section we compute the local divergent part of the quantum effective action
for the purpose of obtaining the beta function. While the computation and the results
are already essentially contained in \cite{MB1}, the calculations here are considerably simplified
and several unnecessary constraints are removed. \par
In particular we compute the beta function in the sector of the $YM$ theory in which
the second parabolic Chern class, $Q$, is negligible
with respect to the quadratic divergence induced by a non-trivial parabolic divisor in the four-dimensional theory.
This is equivalent to require that the reduced topological term, $\frac{Q}{N_2}$, vanishes when $N_2 \rightarrow \infty $,
the reduced $EK$ action being finite on the parabolic divisor. 
In \cite{MB1} it was assumed instead the stronger constraint
that the parabolic Chern class vanishes. \par
Moreover in \cite{MB1} it was assumed that
the structure of the fibration of parabolic Higgs bundles in the four-dimensional theory implies, 
as a consequence of the vanishing of the first and second parabolic Chern classes, the value 
$\frac{2kN(2 \pi)^2}{g^2}$
for the reduced $EK$ classical action 
at leading $\frac{1}{N}$ order, that corresponds to a superposition of $k$ $Z_N$ vortices and
$k$ anti-vortices of lowest vortex number. The anti-vortices occur in \cite{MB1} because a Wilson loop
in the adjoint representation was considered there, that factorizes in the large-$N$ limit
into the product of Wilson loops in the fundamental and in the conjugate representation. \par
Also this constraint is removed, since no four-dimensional structure of the parabolic fibration
survives the $EK$ reduction, because of the vanishing of $\frac{Q}{N_2}$.
Yet, in this section, the correct normalization of the $EK$ classical action, that reproduces the universal
coefficients
of the perturbative the beta function, is obtained 
automatically for vortices of any vortex number.
Finally we find that our construction works directly for a quasi $BPS$ Wilson loop in the fundamental representation
without introducing the adjoint representation, unless we wish to do so. \par
The quantum effective action for a quasi $BPS$ Wilson loop in the fundamental representation
has the following structure:
\bea
\exp({-\Gamma_q})=\int d(zero-modes) \exp{(-\Gamma+Conformal Anomaly)}  
\eea
In the evaluation of the local divergent part of $\Gamma_q$  the global features of the weighted graph
over which $\Gamma_q$ is defined are irrelevant, since the divergences of $\Gamma$, the conformal anomaly and even the zero
modes are locally defined, as we will see momentarily.
$\Gamma$ is the classical effective action in the $ASD$ variables, that includes the non-zero modes of the 
Jacobian of the change of variables. The local part of $\Gamma$ differs by the quantum effective action
on the weighted graph by a conformal 
anomaly, since the  weighted graph is a conformal image of the sphere with the marked divisor over which
$\Gamma$ is defined. \par
Computationally the conformal anomaly plays an important role later, in the calculation of the beta function
for the canonical coupling, but for the moment can be ignored.
$\Gamma_q$  may contain the contribution of zero modes that are not included in  $\Gamma$.
$\Gamma$ contains a globally defined part that arises by the topological term
in the classical $YM$ action. This topological term never contributes to the quantum
effective action in every sector in which it is finite, since in the reduced theory it is divided by
the divergent factor $N_2$. 
This observation makes irrelevant the whole discussion made in \cite{MB1} of the four-dimensional 
constraints that have to be satisfied in order to make the parabolic Chern number vanishing
in a way compatible with localization on vortices. \par
There is a precise relation between $N_2$ and the quadratic divergence at the parabolic points
dictated by the requirement that the cut-off in the longitudinal $z$ plane and in the transverse $u$ plane be equal.
In fact the $EK$ reduction requires the relation $N_2=\frac{H}{2\pi}(\frac{\Lambda_T}{2 \pi })^2$ where
$\frac{\Lambda_T}{2 \pi}=\frac{1}{a_T}$ is the ultraviolet cut-off in the transverse directions of the $EK$ partial reduction.
But it must be $\Lambda_T=\Lambda_L$ to keep the O(4) symmetry of the large-$N$ theory,
that in turn ensures the non-renormalization properties of our quasi $BPS$ Wilson loop. \par
In the reduced $EK$ theory $\Gamma$ is defined on the two-dimensional sphere with marked points,
before mapping conformally to the weighted graph.
Thus the lattice divisor that is the support of the local part of $\Gamma$
is not uniform in general, since this lattice has the same number of points in the interior and in the exterior 
of the Wilson loop. Indeed the weighted graph must have the same number of cusps in the interior and in the
exterior of the Wilson loop.
As a consequence the lattice cut-off differs for each pair of marked points (cusps on the weighted graph)
linked by a backtracking string.
One of these points has a cut-off in the ultraviolet and we call the set of all such points
the ultraviolet divisor. The other one has a cut-off in the infrared and we call
the set of all such points the infrared divisor. \par 
Though $\Gamma$ is the classical action in the $ASD$ variables, $\Gamma$ is not finite
because generically the non-zero modes of the Jacobian determinant of the change of variables plus the
gauge-fixing ghost determinant introduce already
some divergences. These divergences can be computed exactly, being only one loop.
They correct the classical $YM$ action in $\Gamma$ by a $Z^{-1}$ factor that is explicitly displayed below:
\bea
(\frac {N}{2g_W^2}-(2-\frac {1}{3})\frac{N}{(4 \pi)^2} log(\frac {\Lambda}{\tilde \Lambda}))
\sum_{\alpha \ne \beta} \int d^4x
Tr_f (2(\frac{1}{2}F^-_{\alpha \beta})^2) \nonumber \\
=(\frac {N}{2g_W^2}-\frac {5}{3}\frac{N}{(4 \pi)^2} log(\frac {\Lambda}{\tilde \Lambda}))
\sum_{\alpha \ne \beta} \int d^4x Tr_f (2(\frac{1}{2}F^-_{\alpha \beta})^2)\nonumber \\
=\frac {N}{2g_W^2} Z^{-1} \sum_{\alpha \ne \beta} \int d^4x Tr_f (2(\frac{1}{2}F^-_{\alpha \beta})^2)
\eea
where $Z^{-1}$ is given by:
\bea
Z^{-1}=1-\frac{10}{3} \frac{1}{(4 \pi)^2} g_W^2 log(\frac {\Lambda}{\tilde \Lambda})
\eea
and we have added to $g$ the under-script $_W$ to stress that our computation here refers
to the Wilsonean coupling constant.
This formula for $Z$ is actually exact to all orders in the Wilsonean coupling constant, up to finite terms.
The divergent contribution in $\Gamma$ is due entirely to the non-zero modes of the 
localization determinant and of the ghost determinant. The contribution of the holomorphic anomaly
in $\Gamma$ vanishes in any gauge in which $\mu$ can be triangularized. Indeed we can reach this gauge
either in a unitary basis or in a holomorphic basis. The unitary and the holomorphic basis
induce the same Vandermonde determinant of the eigenvalues of $\mu$ in the functional
measure and thus the holomorphic anomaly vanishes in this gauge. \par
$\Gamma$ does not reproduce the perturbative
one-loop beta function. This is due to the peculiar spin term that occurs in the localization determinant
as opposed to the spin term of the gluons determinant in the background field calculation.
The $Z^{-1}$ factor is responsible in the pure $YM$ case of the occurrence
of an anomalous dimension in the formula for the canonical beta function 
as opposed to the $ \cal{N} $ $=1$ $SUSY$ $YM$ case.
The $Z^{-1}$ factor is not present in the $ \cal{N} $ $=1$ $SUSY$ $YM$ theory \cite{MB3} because in that case
the existence of the Nicolai map ensures the cancellation of the non-zero modes
of the determinant of the change of variables versus the gluinos determinant in the light-cone gauge \cite{V,V1}. \par 
Generically there are no normalizable zero modes because the gauge connection and the Higgs
field are both singular at the parabolic points. The resulting pairing 
in the symplectic volume form in function space: 
\bea
\int d^4x Tr(\delta A \wedge \delta A)
\eea
is divergent.
Thus generically in function space and in absence of normalizable zero modes
a less negative beta function than the correct one is obtained.
However there is a special locus of the Higgs field for which normalizable zero modes
exist. This locus corresponds to zeros of the Higgs field, i.e. to vortices equations.
The following volume form, $\omega \wedge  \bar \omega$, is then finite, because the singularity of the
gauge connection is compensated by the zero of the Higgs field:
\bea
\omega =\int d^4x Tr(\delta A_z \wedge \delta D_u)= \frac{2 \pi}{H} \int d^2x Tr(\delta A_z \wedge \delta D_u)
\eea
Since the Higgs field is smooth, for vortices the Hitchin equations of $ASD$ type \cite{ H, S1, S2, S3, S4, S5}
reduce to the case $n_p=0$.
Correspondingly the eigenvalues of the $ASD$ curvature are quantized in such a way
that the local holonomy carries a $Z_N$ charge in the fundamental representation. This quantization follows
by the existence of the zeros of the Higgs field. At the same time the $Z_N$ local holonomy
fixes the normalization of the eigenvalues of the curvature (up to a shift by large gauge transformations
with trivial local holonomy) and thus the value of the action (up to these shifts, see below)
without any extra condition.
For a $Z_N$ vortex of charge $k$ in a $SU(N)$ orbit we get $N-k$ eigenvalues of the curvature equal to
$\frac{2 \pi k}{N}$ and $k$ eigenvalues equal to $\frac{2 \pi (k-N)}{N}$.
The trace of the eigenvalues of the $ASD$ curvature in the fundamental representation is thus:
\bea
(N-k) (\frac{2 \pi k}{N})^2 + k (\frac{2 \pi (k-N)}{N})^2 
=(2 \pi)^2 \frac{k(N-k)}{N}
\eea
In addition each $Z_N$ vortex carries a number of zero modes of the localization determinant
equal to the dimension of the adjoint orbit $ g \lambda g^{-1}$. We do not include zero
modes associated to translations of the vortices since their contribution is
sub-leading in $\frac{1}{N}$.
The complex dimension of an adjoint orbit for a generic parabolic bundle of rank $N$ is given by: 
\bea
dim=\frac{1}{2} (N^2-\sum_i m_i^2)
\eea
where $m_i$ are the multiplicities of the eigenvalues. 
For vortices this reduces to:
\bea
dim=\frac{1}{2} (N^2-k^2-(N-k)^2)=k(N-k)
\eea
 The classical $EK$ reduced action at one point of the ultraviolet divisor on vortices reads:
\bea
\frac{N}{g_W^2}(2 \pi)^2 \frac{k(N-k)}{N}=\frac{8 \pi^2 }{2 g_W^2} k(N-k)
\eea
while at one point of the infrared divisor reads:
\bea
\frac{a^2}{\tilde a^2} \frac{8 \pi^2 }{2 g_W^2} (k(N-k)+ N^2 n^2)
\eea
where the shift $n$ represents the contribution to the eigenvalues of the curvature of a central large gauge transformation
that does not affect the dimension of the vortex adjoint orbit.
Notice that the action at the infrared divisor is suppressed by a power of the ultraviolet cut-off.
We recall that the different cut-off scales arise because there are the same number of cusps internal and external to the
Wilson loop and thus the different cut-off scales are the only way to measure different internal and external areas.
In particular the external area has to go to infinity in the thermodynamic limit.
The existence of two scales seems also necessary in non-perturbative
definitions of the renormalization procedure \cite{Mar}.
The local action at the ultraviolet on vortices is renormalized by the $Z^{-1}$ factor and now also
by the vortices zero modes.
Thus we get for the local part of $\Gamma_q$ at a vortex of charge $k$:
\bea
\exp{(-\Gamma_q(one-vortex))}= \exp{(-\frac{8 \pi^2 }{2 g_W^2} Z^{-1} k(N-k))} (Ha^2)^{-\frac{k(N-k)}{2}} \nonumber \\
\exp{(-\frac{a^2}{\tilde a^2} \frac{8 \pi^2 }{2 g_W^2} (k(N-k)+ N^2 n^2))} 
(H \tilde a^2)^{-\frac{k(N-k)}{2}}
\eea
We omit the contribution of the Vandermonde determinant of the eigenvalues
because it is finite. In addition we do not include it in the canonical effective
action as we did in \cite{MB1} since the canonical normalized action in this paper is not defined by the rescaling of
the eigenvalues of the $ASD$ curvature as in \cite{MB1}. \par
Notice that we have not included a $Z^{-1}$ factor in the contribution to $\Gamma$ at the infrared
because there it is finite and it can be set equal to $1$ by a convenient choice
of the subtraction point.
The factor of $H$ in front of $a^2$
arises from the normalization of the symplectic volume form in the non-commutative theory. 
There is an analogous contribution at the infrared, that is however finite.
$a^{-1}$ is the Pauli-Villars regulator of vortices zero modes at the ultraviolet \cite{Bian}.
The exponent of the Pauli-Villars regulator counts the number of complex
vortices zero modes rather than the number of real zero modes according to our interpretation
of the $D\mu$ integral as a holomorphic integral.
We can avoid the holomorphic counting for the vortices moduli space
by noticing that after continuation to ultra-hyperbolic signature the vortices equations 
are defined on the double cover of $\frac{S^2 \times S^2}{Z_2}$ \cite{Mas}
and that we can use this doubling to define a real pairing of the complex moduli at each location
of the vortices. \par
Alternatively we can consider a Wilson loop in the adjoint representation.
Then we get a factorized contribution from the fundamental and conjugate representation
in the large-$N$ limit.
This implies vortices and anti-vortices in the effective action and a real counting of zero
modes by pairing the vortices with the anti-vortices, as opposed to the holomorphic counting.
Of course the beta function is unchanged because everything gets doubled. \par
The renormalization of the Wilsonean coupling constant now follows immediately
from the effective action at the ultraviolet:
\bea
\frac{8 \pi^2 k(N-k)}{2 g^2_W(\tilde a)}
 = 8 \pi^2 k(N-k)( \frac{1}{2 g^2_W(a)}- \frac{1}{(4\pi)^2} (2+\frac{5}{3})
log (\frac{\tilde a}{a}))
\eea
We should notice that without doing the $EK$ reduction we would obtain the same beta function
once it is observed that the action and the number of zero modes
would have been multiplied by the common factor of $N_2$. \par
It is interesting to write the effective action before the $EK$ reduction since it will be useful
in the computation of the beta function for the canonical coupling.
The unreduced effective action reads:
\bea
\exp{(-\Gamma_q(one-vortex))}= \exp{-( \frac{2 \pi}{H a^2} \frac{8 \pi^2 }{2 g_W^2} Z^{-1} k(N-k))}
\nonumber \\
\exp{-( \frac{2 \pi}{H \tilde a^2} \frac{8 \pi^2 }{2 g_W^2} (k(N-k)+ N^2 n^2) )} \nonumber \\
\exp{( \frac{2 \pi}{H a_T^2} \frac{k(N-k)}{2} log(\frac{1}{H a^2}))} \nonumber \\
\exp{( \frac{2 \pi}{H a_T^2} \frac{k(N-k)}{2} log(\frac{1}{H \tilde a^2}))}
\eea
where the factor of $ \frac{2 \pi}{H a_T^2} $ is the transverse measure over the zero modes two-dimensional sheet.
Indeed, since in the $EK$ partially reduced theory the vortices live on points, in the four-dimensional theory
they live on two-dimensional sheets.
Because of rotational invariance we must have $a=a_T$ and thus the formula for the reduced action follows.
Another way of formulating this condition is that the longitudinal measure on the size of vortices
that we read from the classical action and the transverse measure on the vortices two-dimensional sheet
must coincide.  \par
If a quasi $BPS$ Wilson loop in the adjoint representation is considered as in \cite{MB1} vortices and anti-vortices
contribute to the effective action
and it is possible to pair the holomorphic integral to the anti-holomorphic one. In this case the counting of zero
modes corresponds to the real dimension of the orbit.
For completeness we write the local part of the effective action for a Wilson loop in the adjoint
representation in the notation of \cite{MB1}:
\bea
\exp{(-\Gamma_q)}=  
\prod_p \sum_{k_p,e_p} \exp{-( \frac{2 \pi}{H a^2} \frac{8 \pi^2 }{2 g_W^2} Z^{-1}
k_p(N-k_p)+c.c.)} \nonumber \\
\exp{-( \frac{2 \pi}{H \tilde a^2} \frac{8 \pi^2 }{2 g_W^2} (k_p(N-k_p)+ N^2 e_p^2)+c.c.)} \nonumber \\
\exp{( \frac{2 \pi}{H a_T^2} \frac{k_p(N-k_p)}{2} log(\frac{1}{H a^2})+c.c.)} \nonumber \\
\exp{( \frac{2 \pi}{H a_T^2} \frac{k_p(N-k_p)}{2} log(\frac{1}{H \tilde a^2})+c.c.) }
\eea
where $e_p$ denotes the shift of the local curvature by a central large gauge transformation valued in $Z_N$. \par
The analogous expression for a Wilson loop in the fundamental representation is:
\bea
\exp{(-\Gamma_q)}=  
\prod_p \sum_{k_p,n_p} \exp{-( \frac{2 \pi}{H a^2} \frac{8 \pi^2 }{2 g_W^2} Z^{-1}
k_p(N-k_p))} \nonumber \\
\exp{-( \frac{2 \pi}{H \tilde a^2} \frac{8 \pi^2 }{2 g_W^2} (k_p(N-k_p)+ N^2 n_p^2))} \nonumber \\
\exp{( \frac{2 \pi}{H a_T^2} \frac{k_p(N-k_p)}{2} log(\frac{1}{H a^2}))} \nonumber \\
\exp{( \frac{2 \pi}{H a_T^2} \frac{k_p(N-k_p)}{2} log(\frac{1}{H \tilde a^2})) }
\eea
We now come to the canonical coupling.
We start recalling how the difference
between the Wilsonean and the canonical beta function can be understood
in terms of a rescaling anomaly in the functional integral, in the
$ \cal{N} $ $ = 1$ supersymmetric case, following \cite{AA}. \par
The canonical coupling constant can be related to the Wilsonean one
taking into account an anomalous Jacobian that occurs in the
functional integral: 
\bea
Z= \int \exp(-\frac{N}{2g_{W}^2}S_{YM}(A)) DA \nonumber \\
=\int \exp(-\frac{N}{2g_{W}^2}S_{YM}(g_c A_c)) \frac{D(g_c A_c)}{DA_c} DA_c \nonumber \\
=\int \exp(-\frac{N}{2g_{W}^2}S_{YM}(g_c A_c)+ log \frac{D(g_c A_c)}{DA_c}) DA_c  \nonumber \\
=\int \exp(-\frac{N}{2g_{c}^2}S_{YM}(g_c A_c)) DA_c
\eea
From this relation it follows that:
\bea
\frac{N}{2g_{c}^2} =  \frac{N}{2g_{W}^2} -   S^{-1}_{YM}(g_c A_c) log \frac{D(g_c A_c)}{DA_c}
\eea
We observe that the fields at the ultraviolet and at the infrared are not canonically normalized
in the reduced $EK$ effective action. It is however difficult to canonically normalize the
reduced action because the only possibility would be to rescale 
the vortices eigenvalues. However they are strongly rigid since they are quantized and thus we have not this freedom.
A way out is to use the unreduced effective action.
In the unreduced effective action we can canonically normalize the fields at the ultraviolet and at the infrared by 
writing $H=\frac{Z^{-1}}{g^2_c} H_c$ and $\tilde a^2= Z \tilde a^2_c$ and  by taking fixed these canonically defined scales.
However because of rotational invariance also the canonically defined transverse cut-off must be taken 
fixed $a_T^2= Z a^2_{Tc}$. Finally, we consistently set $a=a_{Tc}$, since the longitudinal measure on the size of vortices
that we read from the classical canonical action and the transverse measure on the vortices two-dimensional sheet
must coincide as in the Wilsonean case.  
This way of defining the canonical transverse cut-off $a_{Tc}$ implies that the rescaled transverse measure does not
depend on the rescaling factor being $g^2_c$ or $g^2_c Z$, as it should be. \par
After a new $EK$ reduction in which this time
$N_2=\frac{2 \pi}{H_c  a^2}$ and the unreduced action is divided by $N_2$ 
we get the following relation between the Wilsonean and the canonical coupling:
\bea
\frac{1}{2 g^2_W}
 = \frac{1}{2 g^2_c}+ \beta_J log g_c + \frac{\beta_J}{4} logZ
\eea
Differentiating this relation
we get for the canonical beta function \cite{MB1}:
\bea
\frac{\partial g_c}{\partial log \Lambda}=\frac{-\beta_0 g_c^3+
\frac {\beta_J}{4} g_c^3 \frac{\partial log Z}{\partial log \Lambda} }{1- \beta_J g_c^2 }
\eea
Now we come to the computation of the anomalous dimension.
Since $Z$ is one-loop exact it leads to the anomalous dimension:
\bea
\frac{\partial log Z}{\partial log a} =-\frac{ \frac{1}{(4\pi)^2} \frac{10}{3} g_W^2}
{1-g_W^2 \frac{1}{(4\pi)^2} \frac{10}{3}
log (\frac{\tilde a}{\sqrt N_D a})}
\eea
where now we have included the contribution of the conformal anomaly, that combines with the subtraction point
to give a finite but arbitrary result for the higher order contributions to the anomalous dimension.
Thus the $RG$ trajectory must be followed along the line $c=-\frac{1}{(4\pi)^2} \frac{10}{3}
 log (\frac{\tilde a}{\sqrt N_D a})=constant$. We observe that it is precisely
the contribution of the conformal anomaly that makes the anomalous dimension a function of the Wilsonean coupling
only, according with the $RG$.
We conclude that the loop equation and the beta function for the quasi $BPS$ Wilson loops
are saturated by the $Z_N$ vortices of the partial $EK$ reduction: 
\bea
F_{z \bar z}-[D_u,D_{\bar u}]= i (\sum_{p} g_p\lambda_pg_p^{-1} \delta^{(2)}(z-z_p)-H1) \nonumber \\
\bar{\partial}_A (D_{u})= 0 \nonumber \\
\partial_{A}(D_{\bar u})=0
\eea

\section{Physical effective charge}

We exploit the residual scheme dependence left in our approach to
find a link with the physical charge between static quark sources in the large-$N$ limit.
The scheme dependence arises because the conformal anomaly in the Wilsonean effective action
is fine-tuned along the $RG$ flow, in order to obtain a finite result for
the higher order contributions to the anomalous dimension.
These contributions are then
finite but contain an arbitrary dimensionless parameter $c$. 
In turn the $c$ dependence of the anomalous dimension determines the $RG$ flow in the
infrared of the canonical coupling. \par
We argue that there should exist in our approach a scheme in which the canonical coupling
coincides with the physical effective charge, essentially because of the very definition
of our canonical coupling via the loop equation. \par
By physical effective charge we mean the coefficient of the Coulomb-like potential
in the complete inter-quark potential, after having separated the linear confining term
proportional to the string tension, by taking second derivatives of the potential
as in \cite{L}:
\bea
V(r)= -\frac{ g^2_{phys}(r)}{4 \pi r}+K r
\eea
where $K$ is the string tension.
With this definition the physical charge may still possess, from a stringy point of view,
non-perturbative string-like contributions \cite{L1} proportional, in the large distance limit,
to inverse powers of $\sqrt K r$. 
In fact it is known from the string model for the large distance inter-quark potential
\cite{L1} that the universal infrared asymptotic value of $\frac{ g^2_{phys}}{4 \pi }$, 
equal to $\frac{\pi}{12}$, might get non-universal corrections in powers of $\sqrt K r$.
These corrections in principle could start at $\frac{1}{\sqrt K r}$ order but
the analysis in \cite{L1} shows that, assuming open/closed string duality,
they in fact could start only at order of $\frac{1}{(\sqrt K r)^2}$.
Another important point is that the beta function of the physical charge thus defined
may possess an infrared  
fixed point without implying that the large-$N$ $YM$  theory be conformal in the infrared,
because the conformal invariance is always
broken by the confining linear term in the inter-quark potential, that shows up in physical Wilson loops. \par
A direct gauge theoretic computation of the physical charge needs 
the evaluation of physical Wilson loops.
In our approach only quasi $BPS$ Wilson loops can be computed directly.
However there are physical Wilson loops that can be obtained by 
analytic continuation from quasi $BPS$ Wilson loops. This will be considered elsewhere \cite{MB4}.
Yet it is natural to assume, because of the universality of the renormalization
procedure, that the canonical coupling that arises in the loop
equation for the quasi $BPS$ Wilson loop coincides in some scheme with the physical charge.
Since the canonical coupling is a local quantity that is determined only
by the local part of the effective action, while the physical charge may get
contributions also from the non-local part, if it does not vanish
in the large-$N$ limit, we can hope to identify the canonical coupling
in some scheme with the effective charge only in the local limit
for the effective action, that is the limit in which the effective action
is actually computed in this paper.\par
Thus we find that if the effective action in the large-$N$ limit implies non-local
interactions between the vortices on which it is localized, then these interactions might
imply the existence of $\frac{1}{\sqrt K r}$ string-like corrections
to the physical effective charge. Otherwise the physical effective charge
is completely accounted by the canonical coupling constant in some scheme. \par
But then this scheme is uniquely determined, as we will see momentarily.
We find that the effective charge has a (non-conformal) infrared fixed point
at the inverse $RG$ invariant scale in the Wilsonean scheme, i.e.
at the Landau pole of the Wilsonean coupling. \par
Given our formulae for the beta function of large-$N$
$YM$: 
\bea
\frac{\partial g_c}{\partial log \Lambda}=\frac{-\beta_0 g_c^3+
\frac {\beta_J}{4} g_c^3 \frac{\partial log Z}{\partial log \Lambda} }{1- \beta_J g_c^2 }
\eea
\bea
\frac{\partial log Z}{\partial log \Lambda} =\frac{ \gamma_0 g_W^2}{1+cg_W^2}
\eea
\bea
\gamma_0=\frac{1}{(4\pi)^2} \frac{10}{3}
\eea
\bea
\frac{\partial g_W}{\partial log \Lambda}=-\beta_0 g_{W}^3
\eea
we inquire as to whether we can find a value of $c$ for which the canonical coupling
coincides with 
the gauge invariant physical charge between static quark sources in the large-$N$ limit. \par
The scheme of the physical charge is uniquely fixed (up perhaps to the mentioned non-local contributions)
by the requirement that the physical charge be continuous and differentiable.
Indeed we can study the behavior of the $RG$ flow as a function of $c$. \par
For $c$ negative the anomalous dimension $ \frac{ \partial log Z}{\partial log \Lambda }$
has a pole at a finite value of $g_W$. Thus the canonical beta function
has a zero in the numerator at a finite distance. However the derivative of the physical charge
does not vanish at that zero. This means that the derivative is discontinuous, since on that point on
the physical charge stays constant. 
Thus $c$ negative is not acceptable in the scheme of the physical charge. \par
For $c$ positive we must distinguish three cases.
For $c > \frac{1}{(4\pi)^2} \frac{\gamma_0}{\beta_0}$ the $RG$ flow from the ultraviolet ends into a cusp,
that is an infrared fixed point at the value $g^2_c=\frac{(4\pi)^2}{4}$, where the beta function has a pole
as in the $ \cal{N} $ $=1$ $SUSY$ $YM$ theory.
However a cusp is not acceptable as the end of the $RG$ flow of the physical charge, because of
the divergence of the derivative of the inter-quark potential. \par
For $\frac{1}{(4\pi)^2} \frac{\gamma_0}{\beta_0} > c $   the flow ends into an infrared fixed point, but the flow is 
continuously differentiable (with zero derivative)
only at the critical value of $c=\frac{1}{(4\pi)^2}  \frac{\gamma_0}{\beta_0}$.
From that point on the canonical coupling remains constant in this scheme. 
The scale at which this occurs is $\Lambda_W^{-1} $, the inverse $RG$ invariant
scale in the Wilsonean scheme. 
We can write the formulae for the beta function in this scheme:
\bea
\frac{\partial g_{phys}}{\partial log \Lambda}=-\beta_0 \frac{g_{phys}^3}{1- \beta_J g_{phys}^2 }
\frac{1}{1+ \frac{1}{(4\pi)^2} \frac{10}{11} g_W^2 } 
\eea
\bea
\frac{\partial g_W}{\partial log \Lambda}=-\beta_0 g_{W}^3
\eea
Eliminating $g_W$ as a function of $\Lambda_W$:
\bea
g_W^2 =-\frac{1}{2 \beta_0 log(r\Lambda_W)}
\eea
we get for the physical beta function as a function of the inter-quark distance:
\bea
\frac{\partial g_{phys}}{\partial log r}=\beta_0 \frac{g_{phys}^3}{1- \beta_J g_{phys}^2 }
\frac{log(r\Lambda_W)}{log(r\Lambda_W)- \frac{15}{121}} 
\eea
It is then easy to see that the junction point with the constant flow is $ C^{\infty}$
and not only continuous and differentiable.
Since $YM$ theory contains only one arbitrary parameter, in our case $\Lambda_W$,
the value of the physical charge at the infrared fixed point is not arbitrary.
However it cannot be found using only
the formula for the beta function, but it needs a direct computation from physical Wilson loops.
It is  possible to perform this computation by analytic continuation
from quasi $BPS$ Wilson loops to physical Wilson loops \cite{MB4} and by our Wilsonean effective
action (the infrared fixed point of the canonical coupling is the Landau pole of the
Wilsonean coupling). \par
Finally we should observe that the prediction of the existence of an infrared fixed point
in the large-$N$ limit for our definition of the physical charge compares favorably,
within the expected order of $\frac{1}{N}$ accuracy,
with the plateau observed in \cite{L} for the group $SU(3)$.
Of course a large-$N$ lattice computation would be more than welcome.

\section{Conclusions}

We have shown the existence in the large-$N$ limit of pure $YM$ and of $ \cal{N} $ $=1$ $SUSY$ $YM$ 
of quasi $BPS$ Wilson loops
that are protected by some of the usual renormalizations, since they have no perimeter and
no cusp divergences for backtracking cusps, as in cases with extended $SUSY$. \par
The existence of such objects is used to localize a version of the 
loop equation in the $ASD$ variables on a saddle-point equation for an effective action.
The proof of localization is obtained by homological methods
as opposed to the cohomological localization in local field theory. \par
The crucial point is that the loop equation reduces to the
insertion of the equation of motion for an effective action
once the local degrees of freedom of the theory
are holographically mapped by a local conformal transformation into backtracking arcs 
ending into cusps at infinity.
Indeed in the new variables and renormalization scheme the Wilsonean renormalized
effective action flows to the ultraviolet by the conformal mapping and, being $AF$, to vanishing coupling. \par
This effective action contains the whole information about the only
remaining renormalization of quasi $BPS$ Wilson loops, i.e. charge renormalization. \par
An explicit formula for the canonical beta function of large-$N$ pure $YM$ theory follows. \par
At the same time this shows the existence of the large-$N$ 
limit of the pure $YM$ theory for this (almost trivial, at least in perturbation theory) class of observables. \par
However one key point is that, since charge renormalization must be the same
in all the sectors of the theory, essentially
by the same token we get information
on the physical charge between static quark sources in the large-$N$ limit. \par
For the future the obvious interesting new game consists in trying to
extend these techniques to those physical Wilson loops that can be obtained
by analytic continuation of quasi $BPS$ Wilson loops. \par
Another interesting aside consists in reproducing the known 
beta function in the $ \cal{N} $ $=1$ $SUSY$ case by the same technique. \par
Finally the present results suggest, in the light of the role
that topological strings play in the localization of the loop equation
on a saddle-point, that a purely stringy approach
to the solution of the loop equation should exist, according to
a long-standing conjecture. In particular the localization on
non-commutative vortices suggests that the equations of motion in target space 
of the string theory dual to large-$N$ $YM$ should be the vortices
equations of $ASD$ type in presence of a $B$ field.

\section{Acknowledgments}

We thank the Galileo Galilei Institute for Theoretical Physics
and the INFN for the hospitality and the financial support during
the completion of this work. \par
We thank the participants to the GGI workshop on "Strongly Coupled Gauge Theories"
for the fruitful atmosphere and in particular Emil Akhmedov, Adriano di Giacomo and Valentin Zakharov 
for several discussions about the existence of quasi $BPS$ Wilson loops. \par
We thank Gabriele Veneziano for pointing out to us that the change of variables
that we employ is actually, in the $\cal{N}$ $=1$ $SUSY$ $YM$ case, the Nicolai map. \par
We thank Martin Luscher for sending us a copy of his October 2004 talk at NBI on "$SU(N)$ Gauge Theories and
the Bosonic String". \par 
We thank the organizers of the meeting of INFN networks on
"Theories of the Fundamental Interactions" (Villa Mondragone, June 2008), Massimo Bianchi,
Loriano Bonora, Luciano Girardello, Kenichi Konishi, for the invitation to speak
about the part of this paper on the $\cal{N}$ $=1$ $SUSY$ $YM$. \par
We thank the organizers of the workshop on "Strings and Strong Interactions: Holography
and beyond" (LNF, September 2008), Stefano Bellucci, Massimo Bianchi, Maria Paola Lombardo,
for the invitation to speak about the part of this paper on the homological localization.

\end{document}